\documentclass{article}
\usepackage{amsfonts}
\usepackage{makeidx}
\usepackage{amsmath}
\usepackage{amssymb}
\usepackage{graphicx}

\newtheorem{theorem}{Theorem}

\input{tcilatex}

\begin{document}

\title{{\Large \textbf{Two dimensional dynamical systems which admit Lie and
Noether symmetries}}}
\author{Michael Tsamparlis\thanks{%
Email: mtsampa@phys.uoa.gr} \ and Andronikos Paliathanasis\thanks{%
Email: anpaliat@phys.uoa.gr} \\
%EndAName
{\small \textit{Faculty of Physics, Department of Astrophysics - Astronomy -
Mechanics,}}\\
{\small \textit{\ University of Athens, Panepistemiopolis, Athens 157 83,
GREECE}}}
\date{}
\maketitle

\begin{abstract}
We consider a dynamical system moving in a Riemannian space and prove two
theorems which relate the Lie point symmetries and the Noether symmetries of
the equation of motion, with the special projective group and the homothetic
group of the space respectively. These theorems are used to classify the two
dimensional Newtonian dynamical systems, which admit Lie point/Noether
symmetries. The results of the study i.e. expressions of forces /
potentials, Lie symmetries, Noether vectors and Noether integrals are
presented in the form of tables for easy reference and convenience. Two
cases are considered, Hamiltonian and non-Hamiltonian systems. The results
are used to determine the Lie / Noether symmetries of two different systems.
The Kepler - Ermakov system, which in general is non-conservative, and the
conservative system with potential similar to the H\`{e}non Heiles
potential. As an additional application, we consider the scalar field
cosmologies inFRW background with no matter, and look for the scalar field
potentials for which the resulting cosmological models are integrable. It is
found that the only integrable scalar field cosmologies are defined by the
exponential and the Unified Dark Matter potential. It is to be noted that in
all aforementioned applications the Lie / Noether symmetry vectors are found
by simply reading the appropriate entry in the relevant tables.
\end{abstract}


PACS - numbers: 2.40.Hw, 4.20.-q, 4.20.Jb, 04.20.Me, 03.20.+i, 02.40.Ky

\section{Introduction}

\label{Introduction}The study of Lie point symmetries of a given system of
ODEs consists of two steps (a)\ the determination of the conditions, which
the components of the Lie symmetry vectors must satisfy and (b) the solution
of the system of these conditions. Step (a) is formal and it is outlined in
e.g. \cite{Olver Book,Stephani book ODES}. These conditions can be quite
involved, but today it is possible to use algebraic computing programs to
derive them. Therefore the essential part of the work is the second step.
For a small number of equations (say up to three) one can possibly employ
again computer algebra to look for a solution of the system. However for a
large number of equations such an attempt is prohibitive and one has to go
back to traditional methods to determine the solution.

The purpose of the present work is to provide an alternative way to solve
the system of Lie point symmetry conditions for the second order equations
of the form:%
\begin{equation}
\ddot{x}^{i}+\Gamma _{jk}^{i}\dot{x}^{j}\dot{x}^{k}=F^{i}.  \label{L2P.1}
\end{equation}%
Here $\Gamma _{jk}^{i}(x^{r})$ are general functions, a dot over a symbol
indicates derivation with respect to the parameter $s$ along the solution
curves and $F^{i}(x^{j})$ is a $C^{\infty }$ vector field. This type of
equations is important, because it contains the equations of motion of a
dynamical system in a Riemannian space, in which the functions $\Gamma
_{jk}^{i}(x^{r})$ are the connection coefficients of the metric, $s$ being
an affine parameter along the trajectory. In the following we assume this
identification of $\Gamma _{jk}^{i}$'s\footnote{%
Of course it is possible to look for a metric for which a given set of $%
\Gamma _{jk}^{i}$ are the connection coefficients, or, even avoid the metric
altogether. However we shall not attempt this in the present work. For such
an attempt see \cite{Fredericks Mahomed Qadir 2007}.}.

The key idea, which is proposed here, is to express the system of Lie
symmetry conditions of (\ref{L2P.1}) in a Riemannian space in terms of
collineation (i.e. symmetry) conditions of the metric. If this is achieved,
then the Lie point symmetries of (\ref{L2P.1}) will be related to the
collineations of the metric, hence their determination  will be transferred
to the geometric problem of determining the generators of a specific type of
collineations of the metric. One then can use of existing results of
Differential Geometry on collineations to produce the solution of the Lie
symmetry problem.

The natural question to ask is: \emph{If the Lie symmetries of the dynamical
systems moving in a given Riemannian space are from the same set of
collineations of the space, how will one select the Lie symmetries of a
specific dynamical system?\ } The answer is as follows. The left hand side
of Equation (\ref{L2P.1}) contains the metric and its derivatives and it is 
\emph{common to all} dynamical systems moving in the same Riemannian space.
Therefore geometry (i.e. collineations)\ enters in the left hand side of (%
\ref{L2P.1}) only. A dynamical system is defined by the force field $F^{i},$
which enters into the right hand side of (\ref{L2P.1}) only. Therefore,
there must exist constraint conditions, which will involve the components of
the collineation vectors and the force field $F^{i}$, which will select the
appropriate Lie symmetries for a specific dynamical system. Indeed Theorem %
\ref{The general conservative system} (see section 3) relates the Lie point
symmetry generators of (\ref{L2P.1}) with the elements of the special
projective Lie algebra of the space where motion occurs, and provides these
necessary constraint conditions. A similar approach can be found in \cite%
{Aminova 2006},\cite{Aminova2010} and for geodesic equations in \cite{Feroze
Mahomed Qadir,Tsamparlis Paliathanasis Lie geodesics Nonlinear Dynamics 2010}%
.

What has been said for the Lie point symmetries of (\ref{L2P.1}) applies
also to Noether symmetries. The Noether symmetries are Lie point symmetries
which satisfy the constraint%
\begin{equation}
X^{\left[ 1\right] }L+L\frac{d\xi }{dt}=\frac{df}{dt}.  \label{L2p.3}
\end{equation}%
Theorem \ref{The Noether symmetries of a conservative system} (see section %
\ref{The Noether symmetries of a conservative dynamical system in a
Riemannian space}) relates the generators of Noether symmetries of (\ref%
{L2P.1}) with the homothetic algebra of the metric and provides the required
constraint conditions.

Before we continue, we recall briefly some definitions from Riemannian
geometry. A\ collineation in a Riemannian space is a vector field $X^{i}$
which satisfies an equation of the form%
\begin{equation}
\mathcal{L}_{X}\mathbf{A=B}  \label{L2p.2}
\end{equation}%
where $\mathcal{L}_{X}$ denotes Lie derivative, $\mathbf{A}$ is a geometric
object (not necessarily a tensor)\ defined in terms of the metric and its
derivatives (e.g. connection coefficients, Ricci tensor, curvature tensor
etc.) and $\mathbf{B}$ is an arbitrary tensor with the same tensor indices
as $\mathbf{A}$. In Table 1 we show examples of collineations which we shall
use in the present work

\begin{center}
.%
\begin{tabular}{lll}
\multicolumn{3}{l}{Table 1: Collineations of space} \\ \hline
\multicolumn{1}{|l}{\textbf{Collineation}} & \multicolumn{1}{|l}{\textbf{\ }$%
\mathbf{A}$} & \multicolumn{1}{|l|}{$\mathbf{B}$} \\ \hline
\multicolumn{1}{|l}{Killing vector (KV)} & \multicolumn{1}{|l}{$g_{ij}$} & 
\multicolumn{1}{|l|}{$0$} \\ \hline
\multicolumn{1}{|l}{Homothetic vector (HV)} & \multicolumn{1}{|l}{$g_{ij}$}
& \multicolumn{1}{|l|}{$\psi g_{ij},~\psi _{,i}=0$} \\ \hline
\multicolumn{1}{|l}{Conformal Killing vector (CKV)} & \multicolumn{1}{|l}{$%
g_{ij}$} & \multicolumn{1}{|l|}{$\psi g_{ij},\psi ,_{i}\neq 0$} \\ \hline
\multicolumn{1}{|l}{Affine Collineation (AC)} & \multicolumn{1}{|l}{$\Gamma
_{jk}^{i}$} & \multicolumn{1}{|l|}{$0$} \\ \hline
\multicolumn{1}{|l}{Projective collineation (PC)} & \multicolumn{1}{|l}{$%
\Gamma _{jk}^{i}$} & \multicolumn{1}{|l|}{$2\phi _{(,j}\delta _{k)}^{i},$ $%
\phi ,_{i}\neq 0$} \\ \hline
\multicolumn{1}{|l}{\small Special Projective collineation (SPC)} & 
\multicolumn{1}{|l}{$\Gamma _{jk}^{i}$} & \multicolumn{1}{|l|}{$2\phi
_{(,j}\delta _{k)}^{i},$ $\phi ,_{i}\neq 0$ and$\phi ,_{jk}=0$} \\ \hline
\end{tabular}
\end{center}

From Differential Geometry we know that the special projective Lie algebra
of the Euclidian space $E^{n}$ consists of the vector fields of Table 2

\begin{center}
.%
\begin{tabular}{l|ll}
\multicolumn{3}{l}{Table 2: Collineations of Euclidean space $E^{n}$} \\ 
\hline
\multicolumn{1}{|l|}{Collineation} & Gradient & \multicolumn{1}{|l|}{
Non-gradient} \\ \hline
\multicolumn{1}{|l|}{Killing vectors (KV)} & $\mathbf{S}_{I}=\delta
_{I}^{i}\partial _{i}$ & \multicolumn{1}{|l|}{$\mathbf{X}_{IJ}=\delta
_{\lbrack I}^{j}\delta _{j]}^{i}x_{j}\partial _{i}$} \\ \hline
\multicolumn{1}{|l|}{Homothetic vector (HV)} & $\mathbf{H}=x^{i}\partial
_{i}~$ & \multicolumn{1}{|l|}{} \\ \hline
\multicolumn{1}{|l|}{Affine Collineation (AC)} & $\mathbf{A}%
_{IJ}=x_{J}\delta _{I}^{i}\partial _{i}~$ & \multicolumn{1}{|l|}{} \\ \hline
\multicolumn{1}{|l|}{Special Projective collineation (SPC)} &  & 
\multicolumn{1}{|l|}{$\mathbf{P}_{I}=S_{I}\mathbf{H}.~$} \\ \hline
\end{tabular}
\end{center}

where the indices $I,J=1,2,\ldots ,n$.

In the following sections we apply Theorem 1 and Theorem 2 to determine all
Newtonian dynamical systems with two degrees of freedom, moving under the
action of a general force $F^{i},$ which admit Lie and\ Noether symmetries.
We also derive for each case the relevant Noether function and the
corresponding Noether integral. The case $F^{i}=0$ corresponds to the Lie
point symmetries of the geodesic equations and has been considered in \cite%
{Tsamparlis Paliathanasis Lie geodesics Nonlinear Dynamics 2010}. The case
of a conservative force has been addressed previously by Sen \cite{Sen} and
more recently by Damianou \emph{et al} \cite{Damianou Sophocleous 1999}$.$
As it will be shown both treatments are incomplete. We demonstrate the use
of the results in two cases. The non-conservative Kepler - Ermakov system 
\cite{Karasu,LeachK,MoyoL} and the case of the H\`{e}non Heiles type
potentials \cite{Leach Henon - Heiles problem,Shrauner}. In both cases we
recover and complete the existing results. Finally we consider the
interesting case of scalar field cosmologies, which is reduced to a system
of two equations of motion in a flat two dimensional Lorentzian space, and
show that the only scalar field potentials which lead to an integrable
cosmological model, are the exponential potential \cite{Russo} and the
Unified Dark Matter (UDM) potential \cite{Bartacca Bartolo
Matarrese,Vasilakos Lukes}.

\section{The Lie point symmetry conditions}

Using the standard approach we derive the Lie point symmetry conditions for
equation (\ref{L2P.1}). We prefer to use the more geometric method outlined
in \cite{Stephani book ODES} rather that the more algebraic (but equivalent)
derivation given e.g. in \cite{Olver Book}.

We write the system of ODEs (\ref{L2P.1}) in the form $\ddot{x}^{i}=\omega
^{i}(x,\dot{x},t)$ where 
\begin{equation}
\omega ^{i}(x,\dot{x},t)=-\Gamma _{jk}^{i}(x)\dot{x}^{j}\dot{x}^{k}-F^{i}.
\label{de.1a}
\end{equation}%
The associated linear operator defined by this system of ODEs is 
\begin{equation}
\mathbf{A}=\frac{\partial }{\partial t}+\dot{x}^{i}\frac{\partial }{\partial
x^{i}}+\omega ^{i}(t,x^{j},\dot{x}^{j})\frac{\partial }{\partial \dot{x}^{i}}%
.  \label{de.2}
\end{equation}%
The condition for a Lie symmetry for the system of equations is \cite%
{Stephani book ODES} 
\begin{equation}
\lbrack \mathbf{X}^{[1]},\mathbf{A}]=\lambda (x^{j})\mathbf{A}  \label{de.3}
\end{equation}%
where $X^{[1]}$ is the first prolongation of the symmetry vector $X=\xi
(t,x)\partial _{t}+\eta ^{i}(t,x)\partial _{x^{i}}$ defined as follows

\begin{equation}
\mathbf{X}^{[1]}=\xi (t,x,\dot{x})\partial _{t}+\eta ^{i}(t,x,\dot{x}%
)\partial _{x^{i}}+G^{[1]i}\partial _{\dot{x}^{i}}.  \label{de.4}
\end{equation}%
$G^{[1]i}$ is the component of the first prolongation $\mathbf{X}^{[1]}$
along $\partial _{\dot{x}^{i}}.$ It is a standard result \cite{Stephani book
ODES} that (\ref{de.3}) leads to the three conditions: 
\begin{align}
-\mathbf{A}\xi & =\lambda  \label{de.5} \\
G^{[1]i}& =\mathbf{A}\eta ^{i}-\dot{x}^{i}\mathbf{A}\xi  \label{de.6} \\
X^{[1]}(\omega ^{i})-\mathbf{A}(G^{[1]{i}})& =-\omega ^{i}\mathbf{A}\xi .
\label{de.7}
\end{align}

For any function $f(t,x^{i}),$ $\mathbf{A}f=df/dt,$ where $%
df/dt=f_{,t}+f_{,i}\dot{x}^{i}$ is the total derivative of $f.$ Using this
result we write the symmetry conditions as%
\begin{align}
\lambda & =-\frac{d\xi }{dt}  \label{de.8} \\
G^{[1]i}& =\frac{d\eta ^{i}}{dt}-\dot{x}^{i}\frac{d\xi }{dt}  \label{de.9} \\
X^{[1]}(\omega ^{i})-A(G^{[1]{i}})& =-\omega ^{i}\frac{d\xi }{dt}.
\label{de.10}
\end{align}

We note that condition (\ref{de.9}) defines the first prolongation $%
G^{[1]i}. $ Condition (\ref{de.8}) gives the factor $\lambda .$ Therefore
the essential condition for a Lie point symmetry is equation (\ref{de.10}).

We introduce the second prolongation $G^{\left[ 2\right] i}$ of $\mathbf{X}$
with the formula%
\begin{equation}
G^{\left[ 2\right] i}\left( t,x^{i},\dot{x}^{i}\right) =\frac{dG^{\left[ 1%
\right] i}}{dt}-\ddot{x}^{i}\frac{d\xi }{dt}=A(G^{[1]{i}})-\omega ^{i}\frac{%
d\xi }{dt}.  \label{de.11}
\end{equation}%
Then the condition for a Lie symmetry becomes \cite{Stephani book ODES}:%
\begin{equation}
G^{\left[ 2\right] i}=X^{[1]}\omega ^{i}.  \label{de3.107}
\end{equation}

After a rather long but standard calculation we find that the Lie point
symmetry conditions of (\ref{L2P.1}) are%
\begin{equation}
L_{\eta }F^{i}+2\xi ,_{t}F^{i}+\eta ^{i},_{tt}=0  \label{PLS.09}
\end{equation}%
\begin{equation}
\left( \xi ,_{k}\delta _{j}^{i}+2\xi ,_{j}\delta _{k}^{i}\right) F^{k}+2\eta
^{i},_{t|j}-\xi ,_{tt}\delta _{j}^{i}=0  \label{PLS.10}
\end{equation}%
\begin{equation}
L_{\eta }\Gamma _{(jk)}^{i}=2\xi ,_{t(j}\delta _{k)}^{i}  \label{PLS.11}
\end{equation}%
\begin{equation}
\xi _{(,i|j}\delta _{r)}^{k}=0.  \label{PLS.12}
\end{equation}

We note that for $F^{i}=0$ we recover the Lie point symmetry conditions of
the geodesic equations (see \cite{Tsamparlis Paliathanasis Lie geodesics
Nonlinear Dynamics 2010})%
\begin{align}
\eta ^{i},_{tt}& =0 \\
2\eta ^{i},_{t|j}-\xi ,_{tt}\delta _{k}^{i}& =0 \\
L_{\eta }\Gamma _{jk}^{i}-2\xi ,_{t(j}\delta _{k)}^{i}& =0 \\
\xi _{(,j|k}\delta _{d)}^{i}& =0.
\end{align}

\section{Lie point symmetries and collineations}

\label{The conservative system}

Following a similar approach as in \cite{Tsamparlis Paliathanasis Lie
geodesics Nonlinear Dynamics 2010} we express the Lie point symmetry
conditions (\ref{PLS.09}) - (\ref{PLS.12}) in terms of the collineations of
the metric.~

Equation (\ref{PLS.12}) means that $\xi _{,j}$ is a gradient Killing vector
(KV) of $g_{ij}.$ This implies that the metric $g_{ij}$ is decomposable.
Equation (\ref{PLS.11}) means that $\eta ^{i}$ is a projective collineation
of the metric with projective function $\xi _{,t}.$ The remaining two
equations are the constraint conditions, which relate the components $\xi
,n^{i}$ of the Lie point symmetry vector with the vector $F^{i}$. Equation (%
\ref{PLS.09}) gives\footnote{$L_{\eta }V_{b}=V_{,bc}\eta ^{c}+\eta
^{c},_{b}V,_{c}$}%
\begin{equation}
\left( L_{\eta }g^{ij}\right) F_{j}+g^{ij}L_{\eta }F_{j}+2\xi
_{,t}g^{ij}F_{j}+\eta _{,tt}^{i}=0.  \label{de.21a}
\end{equation}%
This equation restricts $\eta ^{i}$ further because it relates it directly
to the metric symmetries. Finally equation (\ref{PLS.10}) gives%
\begin{equation}
-\delta _{j}^{i}\xi _{,tt}+\left( \xi _{,j}\delta _{k}^{i}+2\delta
_{j}^{i}\xi _{,k}\right) F^{k}+2\eta _{,tj}^{i}+2\Gamma _{jk}^{i}\eta
_{,t}^{k}=0.  \label{de.21d}
\end{equation}

We conclude that the Lie symmetry equations are equations (\ref{de.21a}) ,(%
\ref{de.21d}) where $\xi (t,x)$ is a gradient KV of the metric $g_{ij}$ and $%
\eta ^{i}\left( t,x\right) $ is a special Projective of the metric $g_{ij}$
with projective function $\xi _{,t}$. We state the solution of the system of
equations (\ref{PLS.09}) - (\ref{PLS.12}) as Theorem \ref{The general
conservative system}.

\begin{theorem}
\label{The general conservative system} The Lie point Symmetries of the
equations of motion of an autonomous system with force function $%
F^{j}(x^{i}),$ $~$in a general Riemannian space with metric $g_{ij},$ namely%
\begin{equation}
\ddot{x}^{i}+\Gamma _{jk}^{i}\dot{x}^{j}\dot{x}^{k}=F^{i}  \label{PP.01}
\end{equation}%
are given in terms of the generators $Y^{i}$ of the special projective Lie
algebra of the metric $g_{ij}$ as follows. \newline
One Lie symmetry vector is $\partial _{t}$ (autonomous equation of motion). 
\newline
Case A: The force is not necessarily conservative

Case A1.

$Y^{i}$ is an element of the Affine algebra of the metric.

The Lie symmetries are:%
\begin{equation}
\mathbf{X}=\left( \frac{1}{2}d_{1}a_{1}t+d_{2}\right) \partial
_{t}+a_{1}Y^{i}\partial _{i},  \label{PP.02}
\end{equation}%
where $a_{1}$ and $d_{1}$ are constants, provided the force satisfies the
condition:%
\begin{equation}
L_{Y}F^{i}+d_{1}F^{i}=0.  \label{PP.03}
\end{equation}%
Case A2.

\qquad $Y^{i}$ is a gradient KV or HV (if it exists)\ and $Y^{i}\neq F^{i}$.

The Lie symmetries are:%
\begin{equation}
\mathbf{X}=~2\psi \int T\left( t\right) dt\partial _{t}+T\left( t\right)
Y^{i}\partial _{i},  \label{PP.04}
\end{equation}%
where the function $T\left( t\right) $ is the solution of the equation%
\begin{equation}
~T_{,tt}=a_{1}T  \label{PP.07}
\end{equation}%
provided the force $F^{i}$ satisfies the condition%
\begin{equation}
L_{\mathbf{Y}}F^{i}+4\psi F^{i}+a_{1}Y^{i}=0.~  \label{PP.05}
\end{equation}

Case A3

$Y^{i}$ is a special PC.

In this case the Lie symmetry vectors are (the index $J$ counts the gradient
KVs)%
\begin{equation}
\mathbf{X}_{J}=\left( C\left( t\right) S_{J}+D\left( t\right) \right)
\partial _{t}+T\left( t\right) Y^{i}\partial _{i},  \label{PP.10}
\end{equation}%
where the functions $C(t),T(t),D(t)$ are solutions of the system of
simultaneous equations%
\begin{equation}
\frac{D_{,t}}{T}=\frac{1}{2}d_{1}~~,~\ \frac{T_{,tt}}{T}=a_{1}~,~\frac{T_{,t}%
}{C}=c_{2}~,~\frac{D_{,tt}}{C}=d_{c}~,~\frac{C_{,t}}{T}=a_{0},  \label{PP.13}
\end{equation}%
provided the force satisfies the conditions 
\begin{eqnarray}
L_{Y}F^{i}+2a_{0}SF^{i}+d_{1}F^{i}-a_{1}Y^{i} &=&0  \label{PP.11} \\
\left( S_{,k}\delta _{j}^{i}+2S,_{j}\delta _{k}^{i}\right) F^{k}-\left(
2Y^{i}{}_{;~j}-a_{0}S\delta _{j}^{i}\right) c_{2}+d_{c}\delta _{j}^{i} &=&0.
\label{PP.12}
\end{eqnarray}

Case B: The force is conservative and $F^{i}=-V^{,i}.$

In addition to the cases of Case A we have the following extra cases.

Case B1.

$Y^{i}$ is a gradient HV and $Y^{i}=\kappa V^{,i},$ where $\kappa $ is a
constant.\newline
In this case the potential is the function generating the gradient HV\ of
the metric and the Lie symmetry vectors are 
\begin{equation}
\mathbf{X}=D\left( t\right) \partial _{t}+T\left( t\right) V^{,i}\partial
_{i}  \label{PP.08}
\end{equation}%
where the functions $D\left( t\right) ,T\left( t\right) $ are the solutions
of the system of simultaneous equations%
\begin{eqnarray}
D_{,tt}-2\psi T_{,t} &=&0  \label{PP.08b} \\
\kappa T_{,tt}+2D_{,t} &=&0.  \label{PP.08c}
\end{eqnarray}%
Case B2.

$Y^{i}$ is a special PC and $Y_{J}^{i}=\lambda S_{J}V^{,i},$ where~$V^{,i}$
is a gradient HV and $S_{J}^{,i}$ is a gradient KV.

The Lie symmetry vectors are:%
\begin{equation*}
X_{J}=\left( C\left( t\right) S_{J}+d_{1}\right) \partial _{t}+T(t)\lambda
S_{J}V^{,i}\partial _{i}
\end{equation*}%
where the functions $C\left( t\right) $ and $T(t)$ are computed from:%
\begin{equation}
\frac{T_{,tt}}{T}+2\frac{C_{,t}}{T}=\lambda _{1}~~,~\frac{T_{,t}}{C}=\lambda
_{2}~,~C_{,t}=a_{0}T  \label{SP.20}
\end{equation}%
provided the potential function satisfies the conditions%
\begin{eqnarray}
L_{\mathbf{Y}_{J}}V^{,i}+\lambda _{1}S_{J}V^{,i} &=&0  \label{SP.21} \\
\left[ \lambda _{2}\left( 2\lambda -a_{0}\right) +\lambda _{1}\right]
S_{J}\delta _{j}^{i}+2(1+\lambda \lambda _{2})S_{J,j}V^{.i} &=&0.
\label{SP.22}
\end{eqnarray}
\end{theorem}

Conditions (\ref{SP.21}) and (\ref{SP.22}) is possible to be written in a
simpler form. Let $H=kV^{,i}$ $(k\neq 0)$ then $Y^{i}=\lambda S_{J}V^{,i}=%
\frac{\lambda }{k}S_{J}H$ $.$

Then condition (\ref{SP.21}) gives (we have changed $\lambda _{1}$ to $%
\lambda _{J})$:%
\begin{equation}
H(\ln S_{J})=\frac{\lambda _{J}}{\lambda }k  \label{SP.21a}
\end{equation}%
which is a condition on the gradient HV $H^{i}.$

Concerning the second condition (\ref{SP.22}) we find:%
\begin{equation}
2(1+\lambda \lambda _{2})(\ln S_{J}),_{j}H^{.i}=-k\left[ \lambda _{2}\left(
2\lambda -a_{0}\right) +\lambda _{1}\right] \delta _{j}^{i}.  \label{SP.21b}
\end{equation}%
This condition is also a constraint between the gradient functions $S_{J},H$
for all $J.$

\section{The Noether symmetries of an autonomous conservative dynamical
system moving in a Riemannian space}

\label{The Noether symmetries of a conservative dynamical system in a
Riemannian space}

Consider a particle moving in the Riemannian space with metric $g_{ij}$
under the influence of the "time" independent potential $V\left(
x^{k}\right) ,$ where by "time"$\ $it is understood the affine parameter
along the trajectory. The Lagrangian of motion is 
\begin{equation}
L=\frac{1}{2}g_{ij}\dot{x}^{i}\dot{x}^{j}-V\left( x^{k}\right) .
\label{NSCS.1}
\end{equation}%
A vector field $\mathbf{X}=\xi \left( t,x^{k}\right) \partial _{t}+\eta
^{i}\left( t,x^{k}\right) \partial _{x^{i}}$ is a Noether Symmetry of the
Lagrangian if condition%
\begin{equation}
\mathbf{X}^{\left[ 1\right] }L+\frac{d\xi }{dt}L=\frac{df}{dt}
\label{NSCS.2}
\end{equation}%
is satisfied, where $\mathbf{X}^{\left[ 1\right] }=\xi \left( t,x^{k}\right)
\partial _{t}+\eta ^{i}\left( t,x^{k}\right) \partial _{x^{i}}+\left( \frac{%
d\eta ^{i}}{dt}-\dot{x}^{i}\frac{d\xi }{dt}\right) \partial _{\dot{x}^{i}}$
is the first prolongation of $\mathbf{X}$ and $f(x^{i},t).$ We compute $%
\mathbf{X}^{\left[ 1\right] }L$ and noting that the resulting equation is an
identity in $\dot{x}^{k},$ we set the coefficient of each power of $\dot{x}%
^{k}$ equal to zero. We find the following conditions for a Noether symmetry
to be admitted by the Lagrangian (\ref{NSCS.1}) 
\begin{eqnarray}
V_{,k}\eta ^{k}+V\xi _{,t} &=&-f_{,t}  \label{NSCS.4} \\
\eta _{,t}^{i}g_{ij}-\xi _{,j}V &=&f_{,j}  \label{NSCS.5} \\
L_{\eta }g_{ij} &=&2\left( \frac{1}{2}\xi _{,t}\right) g_{ij}  \label{NSCS.6}
\\
\xi _{,k} &=&0.  \label{NSCS.7}
\end{eqnarray}

Equation (\ref{NSCS.7}) implies $\xi =\xi \left( t\right) $ and reduces the
system as follows%
\begin{eqnarray}
L_{\eta }g_{ij} &=&2\left( \frac{1}{2}\xi _{,t}\right) g_{ij}  \label{NSCS.8}
\\
V_{,k}\eta ^{k}+V\xi _{,t} &=&-f_{,t}  \label{NSCS.9} \\
\eta _{i,t} &=&f_{,i}.  \label{NSCS.10}
\end{eqnarray}

Equation (\ref{NSCS.8}) implies that $\eta ^{i}$ is a conformal Killing
vector of the metric provided $\xi _{,t}\neq 0.$ Because $g_{ij}$\ is
independent of $t$\ and $\xi =\xi \left( t\right) $\ the $\eta ^{i}$\ must
be is a HV of the metric. This means that $\eta ^{i}\left( t,x\right)
=T\left( t\right) Y^{i}\left( x^{j}\right) $\ where $Y^{i}$\ is a HV. If $%
\xi _{,t}=0$ then $\eta ^{i}$ is a Killing vector of the metric. Equations (%
\ref{NSCS.9}), (\ref{NSCS.10}) are the constraint conditions, which the
Noether symmetry and the potential must satisfy for former to be admitted.

We state the solution of the system of equations (\ref{NSCS.8}) - (\ref%
{NSCS.10}) as Theorem \ref{The Noether symmetries of a conservative system}.

\begin{theorem}
\label{The Noether symmetries of a conservative system}

The Lagrangian:%
\begin{equation}
L=\frac{1}{2}g_{ij}\dot{x}^{i}\dot{x}^{j}-V\left( x^{k}\right)
\label{NSCS.11}
\end{equation}%
of an autonomous conservative dynamical system, moving in a Riemannian space
with metric $g_{ij},$ has two sets of Noether symmetries:

a. The Noether symmetry: 
\begin{equation}
\mathbf{X}=\partial _{t}~,~f=\text{constant}  \label{NSCS.12}
\end{equation}%
which accounts for the autonomous character\footnote{%
Note that autonomous is understood with the meaning ``independent of the
affine parameter along the trajectory''. This parameter need not be the time.%
} of the potential and has the Noether Integral:%
\begin{equation}
E=\frac{1}{2}g_{ij}\dot{x}^{i}\dot{x}^{j}+V\left( x^{k}\right)
\label{NSCS.13}
\end{equation}%
where $E$ is the Hamiltonian of the system.

b. The Noether symmetries which are generated by the homothetic algebra. We
have the following cases. \newline
Case A: The KVs and the HV\ satisfy the condition:%
\begin{equation}
V_{,k}Y^{k}+2\psi _{Y}V+c_{1}=0.  \label{NSCS.14}
\end{equation}%
The Noether symmetry vector is%
\begin{equation}
\mathbf{X}=2\psi _{Y}t\partial _{t}+Y^{i}\partial _{i},~~f=c_{1}t,
\label{NSCS.15}
\end{equation}%
where $T\left( t\right) =a_{0}\neq 0.$ The corresponding Noether integral is
($\psi _{Y}=0$ for a KV\ and $1$ for the HV)%
\begin{equation}
\phi _{II}=2\psi _{Y}tE-g_{ij}Y^{i}\dot{x}^{j}+c_{1}t.  \label{NSCS.16}
\end{equation}%
Case B: The metric admits the gradient KVs $S_{J}$ and the gradient HV $H$
and the potential satisfies the condition%
\begin{equation}
V_{,k}H^{,k}+2\psi _{H}V=c_{2}H+d.  \label{NSCS.17}
\end{equation}%
In this case the Noether symmetry vector and the Noether function are%
\begin{equation}
\mathbf{X}=2\psi _{Y}\int T\left( t\right) dt\partial _{t}+T\left( t\right)
H^{,i}\partial _{i}~~~,~~f\left( t,x^{k}\right) =T_{,t}H\left( x^{k}\right)
~+d\int Tdt.  \label{NSCS.17a}
\end{equation}%
The functions $T(t)$ and $K\left( t\right) $ ($T_{,t}\neq 0)$ are computed
from the relations%
\begin{equation}
T_{,tt}=c_{2}T~,~K_{,t}=d\int Tdt+\text{constant}  \label{NSCS.18}
\end{equation}%
where $c_{2}$ is a constant. The corresponding Noether Integrals are%
\begin{equation}
\phi _{III}\,_{,J}=2\psi _{H}\int T\left( t\right) dt~E-g_{ij}H^{i}\dot{x}%
^{j}+T_{,t}H+d\int Tdt.  \label{NSCS.19}
\end{equation}
\end{theorem}

\section{The Newtonian dynamical systems with two degrees of freedom which
admit Lie symmetries}

\label{Newtonian dynamical systems which admit Lie symmetries}

In this section we apply Theorem \ref{The general conservative system} to
determine all Newtonian dynamical systems with two degrees of freedom which
admit at least one Lie point symmetry. The reason for considering this
problem is that a Lie point symmetry leads to first integrals, which can be
used in many ways to study a given system of differential equations e.g. to
simplify, to determine the integrability of the system etc. Because the
Newtonian systems move in $E^{2}$ we need to consider the generators of the
special projective algebra of $E^{2}$ and then use the constraint conditions
for each case to determine the functional form of the force field $F^{i}.$

We consider Cartesian coordinates so that the metric of the space is: 
\begin{equation}
ds^{2}=dx^{2}+dy^{2}.
\end{equation}%
The special Projective algebra of $E^{2}$ in Cartesian coordinates consists
of the following vector fields (see also Table 2):

\begin{center}
\begin{tabular}{l|ll}
\multicolumn{3}{l}{Table 3: Collineations of Euclidean space $E^{2}$} \\ 
\hline
\multicolumn{1}{|l|}{\textbf{Collineation}} & \textbf{Gradient} & 
\multicolumn{1}{|l|}{\textbf{Non-gradient}} \\ \hline
\multicolumn{1}{|l|}{Killing vectors (KV)} & $\partial _{x}~,~\partial _{y}~$
& \multicolumn{1}{|l|}{$y\partial _{x}-x\partial _{y}$} \\ \hline
\multicolumn{1}{|l|}{Homothetic vector (HV)} & $x\partial _{x}+y\partial
_{y} $ & \multicolumn{1}{|l|}{} \\ \hline
\multicolumn{1}{|l|}{Affine Collineation (AC)} & $x\partial _{x}~,~y\partial
_{y}~,~y\partial _{x}~,~x\partial _{y}$ & \multicolumn{1}{|l|}{$~$} \\ \hline
\multicolumn{1}{|l|}{{\small Special Projective collineation (SPC})} &  & 
\multicolumn{1}{|l|}{$x^{2}\partial _{x}+xy\partial _{y}~,~xy\partial
_{x}+y^{2}\partial _{y}$} \\ \hline
\end{tabular}
\end{center}

We note that the special projective algebra of the two dimensional Lorentz
space 
\begin{equation*}
ds^{2}=-dx^{2}+dy^{2}
\end{equation*}%
is the same with that of the space $E^{2},$ with the difference that the non
gradient Killing vector is replaced with $y\partial _{x}+x\partial _{y}.$ We
shall use this observation in subsection \ref{Scalar field cosmology} where
we study the Lie and Noether symmetries of scalar field cosmologies.

\subsection{ The case of non-conservative forces}

\label{The case of non-conservative forces}We examine first the case where
the force $F^{i}$ is non-conservative. In the next subsection we consider
the case of conservative forces. In certain cases the results are common to
both cases, however for clarity it is better to consider the two cases
separately. Finally for economy of space, easy reference and convenience we
present the results in the form of tables.

In order to indicate how the results of the tables are obtained we consider
case A1 of theorem \ref{The general conservative system}. The Lie point
symmetry vectors for case A1 are given by (\ref{PP.02}) i.e.%
\begin{equation}
\mathbf{X}=\left( \frac{1}{2}d_{1}a_{1}t+d_{2}\right) \partial
_{t}+a_{1}Y^{i}\partial _{i},
\end{equation}%
where $a_{1}$ and $d_{1}$ are constants and $Y^{i}$ is a vector of the
affine algebra of $E^{2}.$ The force field must satisfy condition (\ref%
{PP.03}) i.e.:$L_{Y}\mathbf{F}+d_{1}\mathbf{F}=0.$ Writing $\mathbf{F}%
=F^{x}\partial _{x}~+F^{y}~\partial _{y}$ and $\mathbf{Y}=Y^{x}\partial
_{x}~+Y^{y}~\partial _{y}$ we obtain a system of two differential equations
involving the unknown quantities $F^{x},F^{y}$ and the known quantities $%
Y^{x},Y^{y}.$ For each vector $\mathbf{Y}$ we replace $Y^{x},Y^{y}$ from
Table 3 and solve the system to compute $F^{x},F^{y}.$ For example for the
gradient KV $\partial _{x}$ we have $Y^{x}=1,Y^{y}=0$ and find the solution $%
F^{x}\left( x,y\right) =e^{-dx}f\left( y\right) ,$ $F^{y}\left( x,y\right)
=e^{-dx}g\left( y\right) $ where $d$ is a constant and $f\left( y\right)
,g\left( y\right) $ are arbitrary (but $C^{\infty })$ functions of their
argument. Working similarly we determine the form of the force field for all
cases of Theorem \ref{The general conservative system}. The results are
given in tables 4 and 5.

\begin{center}
\begin{tabular}{lll}
\multicolumn{3}{l}{Table 4: Case A1: The affine algebra} \\ \hline
\multicolumn{1}{|l}{\textbf{Lie }$\downarrow $\textbf{\ }$~~F^{i}\rightarrow 
$} & \multicolumn{1}{|l}{$\mathbf{F}^{x}\left( x,y\right) \mathbf{/F}%
^{\theta }\left( r,\theta \right) $} & \multicolumn{1}{|l|}{$\mathbf{F}%
^{y}\left( x,y\right) \mathbf{/F}^{\theta }\left( r,\theta \right) $} \\ 
\hline
\multicolumn{1}{|l}{$\frac{d}{2}t\partial _{t}+\partial _{x}$} & 
\multicolumn{1}{|l}{$e^{-dx}f\left( y\right) $} & \multicolumn{1}{|l|}{$%
e^{-dx}g\left( y\right) $} \\ \hline
\multicolumn{1}{|l}{$\frac{d}{2}t\partial _{t}+\partial _{y}$} & 
\multicolumn{1}{|l}{$e^{-dy}f\left( x\right) $} & \multicolumn{1}{|l|}{$%
e^{-dy}g\left( x\right) $} \\ \hline
\multicolumn{1}{|l}{$\frac{d}{2}t\partial _{t}+\left( y\partial
_{x}-x\partial _{y}\right) $} & \multicolumn{1}{|l}{$f\left( r\right)
e^{-d\theta }$} & \multicolumn{1}{|l|}{$g\left( r\right) e^{-d\theta }$} \\ 
\hline
\multicolumn{1}{|l}{$\frac{d}{2}t\partial _{t}+x\partial _{x}+y\partial _{y}$%
} & \multicolumn{1}{|l}{$x^{\left( 1-d\right) }f\left( \frac{y}{x}\right) $}
& \multicolumn{1}{|l|}{$x^{\left( 1-d\right) }g\left( \frac{y}{x}\right) $}
\\ \hline
\multicolumn{1}{|l}{$\frac{d}{2}t\partial _{t}+x\partial _{x}$} & 
\multicolumn{1}{|l}{$x^{\left( 1-d\right) }f\left( y\right) $} & 
\multicolumn{1}{|l|}{$x^{-d}g\left( y\right) $} \\ \hline
\multicolumn{1}{|l}{$\frac{d}{2}t\partial _{t}+y\partial _{y}$} & 
\multicolumn{1}{|l}{$y^{-d}f\left( x\right) $} & \multicolumn{1}{|l|}{$%
y^{\left( 1-d\right) }g\left( x\right) $} \\ \hline
\multicolumn{1}{|l}{$\frac{d}{2}t\partial _{t}+y\partial _{x}$} & 
\multicolumn{1}{|l}{$\left( \frac{x}{y}g\left( y\right) +f\left( y\right)
\right) e^{-d\frac{x}{y}}$} & \multicolumn{1}{|l|}{$g\left( y\right) e^{-d%
\frac{x}{y}}$} \\ \hline
\multicolumn{1}{|l}{$\frac{d}{2}t\partial _{t}+x\partial _{y}$} & 
\multicolumn{1}{|l}{$f\left( x\right) e^{-d\frac{y}{x}}$} & 
\multicolumn{1}{|l|}{$\left( \frac{y}{x}f\left( x\right) +g\left( x\right)
\right) e^{-d\frac{y}{x}}$} \\ \hline
&  &  \\ 
\multicolumn{3}{l}{Table 5: Case A2: $Y^{i}$ is a gradient KV or HV and\ $%
T_{,tt}=mT.$} \\ \hline
\multicolumn{1}{|l}{\textbf{Lie }$\downarrow $\textbf{\ }$~~V\rightarrow $}
& \multicolumn{1}{|l}{$\mathbf{F}^{x}\left( x,y\right) $} & 
\multicolumn{1}{|l|}{$\mathbf{F}^{y}\left( x,y\right) $} \\ \hline
\multicolumn{1}{|l}{$T\left( t\right) \partial _{x}$} & \multicolumn{1}{|l}{$%
-mx+f\left( y\right) $} & \multicolumn{1}{|l|}{$g\left( y\right) $} \\ \hline
\multicolumn{1}{|l}{$T\left( t\right) \partial _{y}$} & \multicolumn{1}{|l}{$%
f\left( x\right) $} & \multicolumn{1}{|l|}{$-my+g\left( x\right) $} \\ \hline
\multicolumn{1}{|l}{$2\int T\left( t\right) dt~\partial _{t}+T\left(
t\right) \left( x\partial _{x}+y\partial _{y}\right) $} & 
\multicolumn{1}{|l}{$-\frac{m}{4}x+x^{-3}f\left( \frac{y}{x}\right) $} & 
\multicolumn{1}{|l|}{$-\frac{m}{4}y+y^{-3}g\left( \frac{y}{x}\right) $} \\ 
\hline
\end{tabular}
\end{center}

Case A3: $Y^{i}$ is a special PC

There is only one dynamical system in this case, which is the forced
harmonic oscillator, acted upon the external force $F^{i}=\left( x+a\right)
\partial _{x}+\left( y+b\right) \partial _{y}$. As it can be seen from Table
3 the Lie symmetry algebra is the $sl\left( 4,R\right) .$ This result agrees
with that of \cite{Prince Eliezer}.

Except the above three cases we have to consider the Lie point symmetries
generated from linear combinations of the vectors $Y^{i}.$ It is found that
the only new cases are the ones given given in Table 6 and 7

\begin{center}
\begin{tabular}{lll}
\multicolumn{3}{l}{Table 6: Case A1: $Y^{i}$ is a linear combination of
generators of affine collineations} \\ \hline
\multicolumn{1}{|l}{\textbf{Lie }$\downarrow $\textbf{\ }$~~F^{i}\rightarrow 
$} & \multicolumn{1}{|l}{$\mathbf{F}^{x}\left( x,y\right) \mathbf{/F}%
^{r}\left( r,\theta \right) $} & \multicolumn{1}{|l|}{$\mathbf{F}^{y}\left(
x,y\right) \mathbf{/F}^{\theta }\left( r,\theta \right) $} \\ \hline
\multicolumn{1}{|l}{$\frac{d}{2}t\partial _{t}+\partial _{x}+b\partial _{y}$}
& \multicolumn{1}{|l}{$f\left( y-bx\right) e^{-dx}$} & \multicolumn{1}{|l|}{$%
g\left( y-bx\right) e^{-dx}$} \\ \hline
\multicolumn{1}{|l}{$\frac{d}{2}t\partial _{t}+\left( a+x\right) \partial
_{x}+\left( b+y\right) \partial _{y}$} & \multicolumn{1}{|l}{$f\left( \frac{%
b+y}{a+x}\right) \left( a+x\right) ^{\left( 1-d\right) }$} & 
\multicolumn{1}{|l|}{$g\left( \frac{b+y}{a+x}\right) \left( a+x\right)
^{\left( 1-d\right) }$} \\ \hline
\multicolumn{1}{|l}{$\frac{d}{2}t\partial _{t}+\left( a+x\right) \partial
_{x}+\left( b+hy\right) \partial _{y}$} & \multicolumn{1}{|l}{$f\left(
\left( \frac{b}{h}+y\right) \left( a+bx\right) ^{-\frac{h}{b}}\right) \left(
a+bx\right) ^{1-\frac{d}{b}}$} & \multicolumn{1}{|l|}{$g\left( \left( \frac{b%
}{h}+y\right) \left( a+bx\right) ^{-\frac{h}{b}}\right) \left( a+bx\right) ^{%
\frac{h-d}{b}}$} \\ \hline
\multicolumn{1}{|l}{$\frac{d}{2}t\partial _{t}+\left( x+y\right) \partial
_{x}+\left( x+y\right) \partial _{y}$} & \multicolumn{1}{|l}{$\left( 
\begin{array}{c}
f\left( y-x\right) x+ \\ 
+g\left( y-x\right) 
\end{array}%
\right) \left( y+x\right) ^{-\frac{d}{2}}$} & \multicolumn{1}{|l|}{$\left( 
\begin{array}{c}
f\left( y-x\right) y+ \\ 
-g\left( y-x\right) 
\end{array}%
\right) \left( y+x\right) ^{-\frac{d}{2}}$} \\ \hline
\multicolumn{1}{|l}{$\frac{d}{2}t\partial _{t}+\left( a^{2}x+ay\right)
\partial _{x}$} & \multicolumn{1}{|l}{$a\left( ax+y\right) ^{-\frac{d}{%
1+a^{2}}}\times $} & \multicolumn{1}{|l|}{$a^{2}\left( ax+y\right) ^{-\frac{d%
}{1+a^{2}}}\times \text{\ }$} \\ 
\multicolumn{1}{|l}{$+\left( ax+y\right) \partial _{y}$} & 
\multicolumn{1}{|l}{$\times \left( 
\begin{array}{c}
xa^{2}f\left( y-\frac{x}{a}\right) + \\ 
+g\left( y-\frac{x}{a}\right) 
\end{array}%
\right) $} & \multicolumn{1}{|l|}{$\times \left( 
\begin{array}{c}
af\left( y-\frac{x}{a}\right) + \\ 
-g\left( y-\frac{x}{a}\right) 
\end{array}%
\right) $} \\ \hline
\multicolumn{1}{|l}{$\frac{d}{2}t\partial _{t}+\left( -ay+x\right) \partial
_{x}+\left( ax+y\right) \partial _{y}$} & \multicolumn{1}{|l}{$f\left(
\theta -a\ln r\right) r^{1-d}$} & \multicolumn{1}{|l|}{$g\left( \theta -a\ln
r\right) r^{1-d}$} \\ \hline
&  &  \\ 
\multicolumn{3}{l}{Table 7: Case A2: $Y^{i}$ is a linear combination of
gradient KV or HV and\ $T_{,tt}=mT.$} \\ \hline
\multicolumn{1}{|l}{\textbf{Lie }$\downarrow $\textbf{\ }$F^{i}\rightarrow $}
& \multicolumn{1}{|l}{$\mathbf{F}^{x}\left( x,y\right) $} & 
\multicolumn{1}{|l|}{$\mathbf{F}^{y}\left( x,y\right) $} \\ \hline
\multicolumn{1}{|l}{$T\left( t\right) \left( \partial _{x}+b\partial
_{y}\right) $} & \multicolumn{1}{|l}{$-mx+f\left( y-bx\right) $} & 
\multicolumn{1}{|l|}{$-mbx+g\left( y-bx\right) $} \\ \hline
\multicolumn{1}{|l}{$2\int T\left( t\right) dt~\partial _{t}+$} & 
\multicolumn{1}{|l}{$-\frac{m}{4}\left( a+x\right) +$} & 
\multicolumn{1}{|l|}{$-\frac{m}{4}\left( b+y\right) +$} \\ 
\multicolumn{1}{|l}{$+T\left( t\right) \left[ \left( a+x\right) \partial
_{x}+\left( b+y\right) \partial _{y}\right] $} & \multicolumn{1}{|l}{$%
+f\left( \frac{b+y}{a+x}\right) \left( a+x\right) ^{-3}$} & 
\multicolumn{1}{|l|}{$+g\left( \frac{b+y}{a+x}\right) \left( a+x\right) ^{-3}
$} \\ \hline
\end{tabular}
\end{center}

\subsection{The case of Hamiltonian systems}

\label{The case of Hamiltonian systems}In this subsection we assume $F^{i}$
to be conservative with potential function $V(x,y).$ In this case the
results of the previous Tables differentiate. Furthermore according to
Theorem \ref{The general conservative system} we have to consider the cases
of Case B. The results of the calculations are given below. The results of
the calculations are given in tables 8 and 9.

\begin{center}
\begin{tabular}{llll}
\multicolumn{4}{l}{Table 8: Case A1: The affine algebra} \\ \hline
\multicolumn{1}{|l}{\textbf{Lie }$\downarrow $\textbf{\ }$~~V\rightarrow $}
& \multicolumn{1}{|l}{$\mathbf{d=0}$} & \multicolumn{1}{|l}{$\mathbf{d\neq 0}
$} & \multicolumn{1}{|l|}{$\mathbf{d=2}$} \\ \hline
\multicolumn{1}{|l}{$\frac{d}{2}t\partial _{t}+\partial _{x}$} & 
\multicolumn{1}{|l}{$c_{1}x+f\left( y\right) $} & \multicolumn{1}{|l}{$%
f\left( y\right) e^{-dx}$} & \multicolumn{1}{|l|}{$f\left( y\right) e^{-2x}$}
\\ \hline
\multicolumn{1}{|l}{$\frac{d}{2}t\partial _{t}+\partial _{y}$} & 
\multicolumn{1}{|l}{$c_{1}y+f\left( x\right) $} & \multicolumn{1}{|l}{$%
f\left( x\right) e^{-dy}$} & \multicolumn{1}{|l|}{$f\left( x\right) e^{-2y}$}
\\ \hline
\multicolumn{1}{|l}{$\frac{d}{2}t\partial _{t}+\left( y\partial
_{x}-x\partial _{y}\right) $} & \multicolumn{1}{|l}{$\theta +f\left(
r\right) $} & \multicolumn{1}{|l}{$f\left( r\right) e^{-d\theta }$} & 
\multicolumn{1}{|l|}{$f\left( r\right) e^{-2\theta }$} \\ \hline
\multicolumn{1}{|l}{$\frac{d}{2}t\partial _{t}+\left( x\partial
_{x}+y\partial _{y}\right) $} & \multicolumn{1}{|l}{$x^{2}f\left( \frac{y}{x}%
\right) $} & \multicolumn{1}{|l}{$x^{2-d}f\left( \frac{y}{x}\right) $} & 
\multicolumn{1}{|l|}{$c_{1}\ln x~+f\left( \frac{y}{x}\right) $} \\ \hline
\multicolumn{1}{|l}{$\frac{d}{2}t\partial _{t}+x\partial _{x}$} & 
\multicolumn{1}{|l}{$c_{1}x^{2}+f\left( y\right) $} & \multicolumn{1}{|l}{$%
\nexists $} & \multicolumn{1}{|l|}{$\nexists $} \\ \hline
\multicolumn{1}{|l}{$\frac{d}{2}t\partial _{t}+y\partial _{y}$} & 
\multicolumn{1}{|l}{$c_{1}y^{2}+f\left( x\right) $} & \multicolumn{1}{|l}{$%
\nexists $} & \multicolumn{1}{|l|}{$\nexists $} \\ \hline
\multicolumn{1}{|l}{$\frac{d}{2}t\partial _{t}+y\partial _{x}$} & 
\multicolumn{1}{|l}{$x^{2}+y^{2}+c_{1}x$} & \multicolumn{1}{|l}{$\nexists $}
& \multicolumn{1}{|l|}{$\nexists $} \\ \hline
\multicolumn{1}{|l}{$\frac{d}{2}t\partial _{t}+x\partial _{y}$} & 
\multicolumn{1}{|l}{$x^{2}+y^{2}+c_{1}y$} & \multicolumn{1}{|l}{$\nexists $}
& \multicolumn{1}{|l|}{$\nexists $} \\ \hline
&  &  &  \\ 
\multicolumn{4}{l}{Table 9: Case A2: $Y^{i}$ is a gradient KV or HV} \\ 
\hline
\multicolumn{1}{|l}{\textbf{Lie }$\downarrow $\textbf{\ }$~~V\rightarrow $}
& \multicolumn{3}{|l|}{$\mathbf{T}_{,tt}\mathbf{=mT.}$} \\ \hline
\multicolumn{1}{|l}{$T\left( t\right) \partial _{x}$} & \multicolumn{3}{|l|}{%
$-\frac{mx^{2}}{2}+c_{1}x+f\left( y\right) $} \\ \hline
\multicolumn{1}{|l}{$T\left( t\right) \partial _{y}$} & \multicolumn{3}{|l|}{%
$-\frac{my^{2}}{2}+c_{1}y+f\left( x\right) $} \\ \hline
\multicolumn{1}{|l}{$2\int T\left( t\right) dt~\partial _{t}+T\left(
t\right) \left( x\partial _{x}+y\partial _{y}\right) $} & 
\multicolumn{3}{|l|}{$-\frac{m}{8}\left( x^{2}+y^{2}\right) +\frac{1}{x^{2}}%
f\left( \frac{y}{x}\right) $} \\ \hline
\end{tabular}
\end{center}

The linear combinations give the following new cases listed in tables 10,11
and 12

\begin{center}
\begin{tabular}{lll}
\multicolumn{3}{l}{Table 10: Case A1: $Y^{i}$ is a linear combination
generators of ACs} \\ \hline
\multicolumn{1}{|l}{\textbf{Lie }$\downarrow $\textbf{\ }$~~V\rightarrow $}
& \multicolumn{1}{|l}{$\mathbf{d=0}$} & \multicolumn{1}{|l|}{$\mathbf{d\neq 0%
}$} \\ \hline
\multicolumn{1}{|l}{$\frac{d}{2}t\partial _{t}+a\partial _{x}+b\partial _{y}$%
} & \multicolumn{1}{|l}{$f\left( ay-bx\right) $} & \multicolumn{1}{|l|}{$%
\left[ c_{1}+f\left( ay-bx\right) \right] e^{-d\frac{x}{a}}$} \\ \hline
\multicolumn{1}{|l}{$\frac{d}{2}t\partial _{t}+\left( a+x\right) \partial
_{x}+\left( b+y\right) \partial _{y}$} & \multicolumn{1}{|l}{$f\left( \frac{%
b+y}{a+x}\right) \left( a+x\right) ^{2}$} & \multicolumn{1}{|l|}{$f\left( 
\frac{b+y}{a+x}\right) \left( a+x\right) ^{\left( 2-d\right) }$} \\ \hline
\multicolumn{1}{|l}{$\frac{d}{2}t\partial _{t}+\left( x+y\right) \partial
_{x}+\left( x+y\right) \partial _{y}$} & \multicolumn{1}{|l}{$f\left(
y-x\right) +c_{1}\left( x+y\right) ^{2}$} & \multicolumn{1}{|l|}{$\left(
x+y\right) ^{\left( 2-\frac{d}{2}\right) ~~}$} \\ \hline
\multicolumn{1}{|l}{$\frac{d}{2}t\partial _{t}+\left( a^{2}x+ay\right)
\partial _{x}+\left( ax+y\right) \partial _{y}$} & \multicolumn{1}{|l}{$%
c_{1}\left( x^{2}+y^{2}\right) ~+f\left( ay-x\right) $} & 
\multicolumn{1}{|l|}{$c_{1}\left( ax+y\right) ^{\left( 2-\frac{d}{1+a^{2}}%
\right) }\text{\ }$} \\ \hline
\multicolumn{1}{|l}{$\frac{d}{2}t\partial _{t}+\left( -ay+x\right) \partial
_{x}+\left( ax+y\right) \partial _{y}$} & \multicolumn{1}{|l}{$f\left(
\theta -a\ln r\right) r^{2}$} & \multicolumn{1}{|l|}{$f\left( \theta -a\ln
r\right) r^{2-d}$} \\ \hline
&  &  \\ 
\multicolumn{3}{l}{Table 11: Case A1\textbf{\ }(continuation of Table 10)}
\\ \hline
\multicolumn{1}{|l}{\textbf{Lie }$\downarrow $\textbf{\ }$~~V\rightarrow $}
& \multicolumn{1}{|l}{$\mathbf{d=2}$} & \multicolumn{1}{|l|}{$\mathbf{d=1}$}
\\ \hline
\multicolumn{1}{|l}{$\frac{d}{2}t\partial _{t}+\partial _{x}+b\partial _{y}$}
& \multicolumn{1}{|l}{$\left[ c_{1}+f\left( y-bx\right) \right] e^{-2\frac{x%
}{a}}$} & \multicolumn{1}{|l|}{$\left[ c_{1}+f\left( y-bx\right) \right] e^{-%
\frac{x}{a}}$} \\ \hline
\multicolumn{1}{|l}{$\frac{d}{2}t\partial _{t}+\left( a+x\right) \partial
_{x}+\left( b+y\right) \partial _{y}$} & \multicolumn{1}{|l}{$f\left( \frac{%
b+y}{a+x}\right) +c_{1}\ln \left( a+x\right) $} & \multicolumn{1}{|l|}{$%
f\left( \frac{b+y}{a+x}\right) \left( a+x\right) $} \\ \hline
\multicolumn{1}{|l}{$\frac{d}{2}t\partial _{t}+\left( x+y\right) \partial
_{x}+\left( x+y\right) \partial _{y}$} & \multicolumn{1}{|l}{$\left(
x+y\right) $} & \multicolumn{1}{|l|}{$\left( x+y\right) ^{\frac{3}{2}}$} \\ 
\hline
\multicolumn{1}{|l}{$\frac{d}{2}t\partial _{t}+\left( a^{2}x+ay\right)
\partial _{x}+\left( ax+y\right) \partial _{y}$} & \multicolumn{1}{|l}{$\ln
\left( ax+y\right) \text{\ ~}\left( d=2\left( 1+a^{2}\right) \right) $} & 
\multicolumn{1}{|l|}{$\left( ax+y\right) ^{\left( \frac{1+2a^{2}}{1+a^{2}}%
\right) }\text{\ }$} \\ \hline
\multicolumn{1}{|l}{$\frac{d}{2}t\partial _{t}+\left( -ay+x\right) \partial
_{x}+\left( ax+y\right) \partial _{y}$} & \multicolumn{1}{|l}{$c_{1}\ln
r+f\left( \theta -a\ln r\right) $} & \multicolumn{1}{|l|}{$c_{1}r+f\left(
\theta -a\ln r\right) r$} \\ \hline
\multicolumn{3}{l}{} \\ 
\multicolumn{3}{l}{Table 12: Case A2 $Y^{i}$ is a linear combination
generators of the gradient HVs} \\ \hline
\multicolumn{1}{|l}{\textbf{Lie }$\downarrow $\textbf{\ }$V\rightarrow $} & 
\multicolumn{2}{|l|}{$\mathbf{T}_{,tt}\mathbf{=mT}$} \\ \hline
\multicolumn{1}{|l}{$T\left( t\right) \left( a\partial _{x}+b\partial
_{y}\right) $} & \multicolumn{2}{|l|}{$-\frac{m}{2}(x^{2}+y^{2})+c_{1}x+f%
\left( ay-bx\right) $} \\ \hline
\multicolumn{1}{|l}{$2\int T\left( t\right) dt~\partial _{t}+T\left(
t\right) \left[ \left( a+x\right) \partial _{x}+\left( b+y\right) \partial
_{y}\right] $} & \multicolumn{2}{|l|}{$-\frac{m}{8}\left(
x^{2}+y^{2}+2ax+2by\right) +\left( a+x\right) ^{-2}f\left( \frac{b+y}{a+x}%
\right) $} \\ \hline
\end{tabular}
\bigskip 
\end{center}

As it was stated in section \ref{Introduction} the determination of all two
dimensional potentials which admit a Lie point symmetry has been addressed
previously in \cite{Sen} and \cite{Damianou Sophocleous 1999}. Our results
contain the results of both these papers and additionally some cases
missing, mainly in the linear combinations of the HV with the KVs. Obviously
the derivation discussed above is systematic and can be generalized to
higher dimensions in a straightforward manner. However the possible
symmetries increase dramatically so this task can only be considered
elsewhere.

\section{The Newtonian dynamical systems with two degrees of freedom which
admit Noether symmetries}

\label{Newtonian Hamiltomian systems which admit Noether Symmetries}

Noether symmetries are associated with a Lagrangian. Therefore we consider
only the case in which the force $F^{i}$ is conservative. Furthermore,
Noether symmetries are special Lie point symmetries, hence we look into the
two dimensional potentials which admit a Lie point symmetry. These
potentials were determined in the previous section. We apply Theorem \ref%
{The Noether symmetries of a conservative system} to these potentials and
select the potentials which admit a Noether symmetry. The calculations are
similar to the ones for the Lie point symmetries and are omitted. The
results are listed in tables 13,14,15 and 16.

\begin{center}
\ 
\begin{tabular}{ll}
\multicolumn{2}{l}{Table 13: Case A. $Y^{i}$ is a HV} \\ \hline
\multicolumn{1}{|l}{\textbf{Noether Symmetry}} & \multicolumn{1}{|l|}{$%
\mathbf{V}\left( x,y\right) $} \\ \hline
\multicolumn{1}{|l}{$\partial _{x}$} & \multicolumn{1}{|l|}{$cx+f\left(
y\right) $} \\ \hline
\multicolumn{1}{|l}{$\partial _{y}$} & \multicolumn{1}{|l|}{$cy+f\left(
x\right) $} \\ \hline
\multicolumn{1}{|l}{$y\partial _{x}-x\partial _{y}$} & \multicolumn{1}{|l|}{$%
c\theta +f\left( r\right) $} \\ \hline
\multicolumn{1}{|l}{$2t\partial _{t}+x\partial _{x}+y\partial _{y}$} & 
\multicolumn{1}{|l|}{$x^{-2}f\left( \frac{y}{x}\right) $} \\ \hline
&  \\ 
\multicolumn{2}{l}{Table 14: Case A. $Y^{i}$ is a linear combination of HVs}
\\ \hline
\multicolumn{1}{|l}{\textbf{Noether Symmetries}} & \multicolumn{1}{|l|}{$%
\mathbf{V}\left( x,y\right) $} \\ \hline
\multicolumn{1}{|l}{$\partial _{x}+b\partial _{y}$} & \multicolumn{1}{|l|}{$%
f\left( y-bx\right) -cx$} \\ \hline
\multicolumn{1}{|l}{$\left( a+y\right) \partial _{x}+\left( b-x\right)
\partial _{y}$} & \multicolumn{1}{|l|}{$f\left( \frac{1}{2}\left(
x^{2}+y^{2}\right) +ay-bx\right) $} \\ \hline
\multicolumn{1}{|l}{$2t\partial _{t}+\left( x+ay\right) \partial _{x}+\left(
y-ax\right) \partial _{y}$} & \multicolumn{1}{|l|}{$r^{-2}~f\left( \theta
-a\ln r\right) $} \\ \hline
\multicolumn{1}{|l}{$2t\partial _{t}+\left( a+x\right) \partial _{x}+\left(
b+y\right) \partial _{y}$} & \multicolumn{1}{|l|}{$f\left( \frac{b+x}{a+x}%
\right) \left( a+x\right) ^{-2}-c\left( a+x\right) ^{-2}\left( \frac{1}{2}%
x^{2}+ax\right) $} \\ \hline
\end{tabular}

\begin{tabular}{ll}
\multicolumn{2}{l}{Table 15: Case B. $Y^{i}$ is a gradient HV} \\ \hline
\multicolumn{1}{|l}{\textbf{Noether \ }$\downarrow $\textbf{\ }$V\rightarrow 
$} & \multicolumn{1}{|l|}{$\mathbf{T}_{,tt}\mathbf{=mT}$} \\ \hline
\multicolumn{1}{|l}{$\,T\left( t\right) \partial _{x}$} & 
\multicolumn{1}{|l|}{$f\left( y\right) -cx-\frac{m}{2}x^{2}$} \\ \hline
\multicolumn{1}{|l}{$T\left( t\right) \partial _{y}$} & \multicolumn{1}{|l|}{%
$f\left( x\right) -cy-\frac{m}{2}y^{2}$} \\ \hline
\multicolumn{1}{|l}{$2\int T\left( t\right) dt~\partial _{t}+T\left(
t\right) \left( x\partial _{x}+y\partial _{y}\right) $} & 
\multicolumn{1}{|l|}{$x^{-2}f\left( \frac{y}{x}\right) -\frac{m}{8}\left(
x^{2}+y^{2}\right) $} \\ \hline
&  \\ 
\multicolumn{2}{l}{Table 16: Case B. $Y^{i}$ is a linear combination of
gradient HVs} \\ \hline
\multicolumn{1}{|l}{\textbf{Noether }$\downarrow $\textbf{\ }$~~V\rightarrow 
$} & \multicolumn{1}{|l|}{$\mathbf{T}_{,tt}\mathbf{=mT}$} \\ \hline
\multicolumn{1}{|l}{$\,T\left( t\right) \partial _{x}+bT\left( t\right)
\partial _{y}$} & \multicolumn{1}{|l|}{$-\frac{m}{2}\left(
x^{2}+y^{2}\right) -\frac{m}{2}(y-bx)^{2}+f\left( y-bx\right) -cx$} \\ \hline
\multicolumn{1}{|l}{$2\int T\left( t\right) dt~\partial _{t}+T\left(
t\right) \left( \left( a+x\right) \partial _{x}+\left( b+y\right) \partial
_{y}\right) $} & \multicolumn{1}{|l|}{$f\left( \frac{b+x}{a+x}\right) \left(
a+x\right) ^{-2}-\frac{c}{2}\left( a+x\right) ^{-2}x\left( x+2a\right) $} \\ 
\multicolumn{1}{|l}{} & \multicolumn{1}{|l|}{$-\frac{xm\left( x+2a\right) }{%
8\left( a+x\right) ^{4}}\left\{ 
\begin{array}{c}
\left( \left( a+x\right) ^{2}+a^{2}\right) y\left( y+2b\right) + \\ 
+x\left( x+2a\right) \left( b+\left( a+x\right) \right) \left( -b+\left(
a+x\right) \right) 
\end{array}%
\right\} $} \\ \hline
\end{tabular}
\\[0pt]
\end{center}

\section{Applications}

In this section we demonstrate the application of the results of sections %
\ref{Newtonian dynamical systems which admit Lie symmetries} and \ref%
{Newtonian Hamiltomian systems which admit Noether Symmetries} in three
cases. The first case is the Kepler-Ermakov system, which (in general) is
not a conservative dynamical system, the second is the H\`{e}non - Heiles
type potential and the third is the scalar field cosmology.

\subsection{The Lie Symmetries of the Kepler-Ermakov system.}

The Ermakov systems are time dependent dynamical systems, which contain an
arbitrary function of time (the frequency function) and two arbitrary
homogeneous functions of dynamical variables. A central feature of Ermakov
systems is their property of always having a first integral.  The
Kepler-Ermakov system is an autonomous Ermakov system defined by the
equations \cite{Athorne 1991} 
\begin{eqnarray}
\ddot{x}+\frac{x}{r^{3}}H\left( x,y\right) -\frac{1}{x^{3}}f\left( \frac{y}{x%
}\right) &=&0  \label{Kepler-Ermakov 1} \\
\ddot{y}+\frac{y}{r^{3}}H\left( x,y\right) -\frac{1}{y^{3}}g\left( \frac{y}{x%
}\right) &=&0  \label{Kepler-Ermakov 2}
\end{eqnarray}%
where $H,f,g~$are arbitrary functions. In \cite{Karasu} it has been shown
that this system admits Lie point symmetries for certain forms of the
function $H\left( x,y\right) .$ Furthermore it has been shown that for
special classes of these equations there exists a Lagrangian (see also \cite%
{LeachK}).

In the following we demonstrate the use of our results by finding the Lie
symmetries simply by reading the entries of the proper tables. Looking at
the tables we find that equations (\ref{Kepler-Ermakov 1}), (\ref%
{Kepler-Ermakov 2}) admit a Lie point symmetry for the following two cases.%
\newline
Case 1.

When $H\left( x,y\right) =\frac{h\left( \frac{y}{x}\right) }{x}.$ Then line
4 of Table 4 applies for $d=4.$ Also line 3 of Table 5 for $m=0.$ Therefore
the Lie point symmetries are: 
\begin{equation}
X=\left( c_{1}+c_{2}2t+c_{3}t^{2}\right) \partial _{t}+\left(
c_{2}x+c_{3}tx\right) \partial _{x}+\left( c_{2}y+c_{3}ty\right) \partial
_{y}.  \label{NEP.01}
\end{equation}%
Case 2.

When $H\left( x,y\right) =\omega ^{2}r^{3}+\frac{h\left( \frac{y}{x}\right) 
}{x}$ where $m=-4\omega ^{2}$ and $m\neq 0$. In this case line 3 of Table 5
for $m\neq 0$ applies and the Lie point symmetry generator is 
\begin{equation*}
X=\left( c_{1}-\frac{c_{2}}{\omega }\cos \left( 2\omega t\right) +\frac{c_{3}%
}{\omega }\sin \left( 2\omega t\right) \right) \partial _{t}+\left(
c_{2}\sin \left( 2\omega t\right) +c_{3}\cos \left( 2\omega t\right) \right)
x\partial _{x}+\left( c_{2}\sin \left( 2\omega t\right) +c_{3}\cos \left(
2\omega t\right) \right) y\partial _{y}.
\end{equation*}%
These symmetries coincide with the ones found in \cite{Karasu}. We note that
in both cases the Lie symmetry vectors come from the HV $x\partial
_{x}+y\partial _{y}$ of the Euclidean metric.

In a subsequent publication \cite{LeachK} it was shown that the Lagrangian
considered in \cite{Karasu} was incorrect and that the correct Lagrangian
is: 
\begin{equation}
L=\frac{1}{2}\left( \dot{r}^{2}+r^{2}\dot{\theta}^{2}\right) -\frac{1}{2}%
\omega ^{2}r^{2}-\frac{\mu }{2r^{2}}-\frac{C\left( \theta \right) }{2r^{2}}
\end{equation}
where $C(\theta )=\sec ^{2}\theta f(\tan \theta )+\csc ^{2}\theta g(\tan
\theta )$ and the functions $f,g$ satisfy two compatibility conditions (see
equation (5.2) of \cite{LeachK}).

In order to find the Noether symmetries of this Lagrangian we note that for
every function $C\left( \theta \right) $ the potential is of the form $%
V\left( r,\theta \right) =\frac{m}{2}\left( x^{2}+y^{2}\right)
+x^{-2}V\left( \frac{y}{x}\right) .$ This means that line 4 of Table 13 and
line 3 of Table 15 with $\omega =0~,m=0$ apply. It follows that the Lie
symmetries are also Noether symmetries and that the Noether Integrals (in
addition to the Hamiltonian $E$) corresponding the these Noether symmetries
are 
\begin{eqnarray}
I_{1} &=&2tE-r\dot{r} \\
I_{2} &=&t^{2}E-tr\dot{r}+\frac{1}{2}r^{2}.
\end{eqnarray}

In total we have three Noether integrals\footnote{%
When $\omega \neq 0$ only line 3 of Table 15 for $m=-4\omega ^{2}$ applies
and the Lie symmetry is also Noether symmetry with Noether integrals 
\begin{eqnarray}
I_{1}^{\prime } &=&-\frac{1}{\omega }\cos \left( 2\omega t\right) E-\sin
\left( 2\omega t\right) r\dot{r}+\omega \cos \left( 2\omega t\right) r^{2} \\
I_{2}^{^{\prime }} &=&\frac{1}{\omega }\sin \left( 2\omega t\right) E-\cos
\left( 2\omega t\right) r\dot{r}-\omega \sin \left( 2\omega t\right) r^{2}.
\end{eqnarray}%
}. Since we do not look for generalized symmetries, we do not expect to find
the Ermakov - Lewis invariant \cite{MoyoL}.

\subsection{On the H\`{e}non - Heiles potential}

The H\`{e}non - Heiles potential 
\begin{equation*}
V\left( x,y\right) =\frac{1}{2}\left( x^{2}+y^{2}\right) +x^{2}y-\frac{1}{3}%
y^{2}
\end{equation*}%
has been used as a model for the galactic cluster. Computer analysis has
suggested that for sufficiently small values of the energy, there exists a
first integral independent of energy. In \cite{Leach Henon - Heiles problem}
it is proposed to study if there exists a Lie point symmetry of the
potential which could justify such a first integral. Working in a slightly
more general scenario, in \cite{Leach Henon - Heiles problem} are considered
potentials of the form 
\begin{equation}
V\left( x,y\right) =\frac{1}{2}\left( x^{2}+y^{2}\right)
+Ax^{3}+Bx^{2}y+Cxy^{2}+Dy^{3}  \label{Leach1}
\end{equation}%
where \thinspace $A,B,C,D$ are real parameters. The H\`{e}non - Heiles
potential is the special case for \thinspace $A=C=0,B=1,D=-\frac{1}{3}$.

Using standard Lie analysis in \cite{Leach Henon - Heiles problem} it is
shown that only the potentials $V_{1}\left( x,y\right) =\frac{1}{2}\left(
x^{2}+y^{2}\right) +x^{3},$ \ $V_{2}\left( x,y\right) =\frac{1}{2}\left(
x^{2}+y^{2}\right) +y^{3},$ \ $V_{3}\left( x,y\right) =\frac{1}{2}\left(
x^{2}+y^{2}\right) \pm \left( ay\pm x\right) ^{3},$ \ $V_{4}\left(
x,y\right) =\frac{1}{2}\left( x^{2}+y^{2}\right) \pm \left( ay\mp x\right)
^{3}$ admit Lie point symmetries, hence the H\`{e}non - Heiles potential
does not admit a Lie symmetry and the existence of a first integral it is
not justified. We apply the results of sections \ref{Newtonian dynamical
systems which admit Lie symmetries}, \ref{Newtonian Hamiltomian systems
which admit Noether Symmetries} to give the Lie point symmetries and the
Noether quantities of these potentials, simply by reading the relevant
tables.

The potential $V_{1}\left( x,y\right) $ is of the form $cy^{2}+f\left(
x\right) $ and belongs to the types:

\begin{center}
\begin{tabular}{|l|l|l|l|l|}
\hline
\textbf{Table} & \textbf{Line} & \textbf{Parameters} & \textbf{Lie } & 
\textbf{Type of vector} \\ \hline
8 & 6 & $~d=0,c_{1}=\frac{1}{2},f\left( x\right) =\frac{1}{2}x^{2}+x^{3}$ & $%
y\partial _{y}$ & Affine \\ \hline
9 & 2 & $m=-1,c_{1}=0,f\left( x\right) =\frac{1}{2}x^{2}+x^{3}$ & ($\cos t$
or $\sin t)\partial _{y}$ & gradient KV \\ \hline
\end{tabular}
\end{center}

Therefore the Lie symmetries admitted by this potential are: 
\begin{equation*}
X=c_{0}\partial _{t}+c_{1}\sin t\partial _{y}\ +c_{2}\cos t\partial
_{y}+c_{3}y\partial _{y}.
\end{equation*}

We note that the Lie symmetry $y\partial _{y},$ which is due to the Affine
collineation, has not been found in \cite{Leach Henon - Heiles problem}.

The potential $V_{2}\left( x,y\right) $\textbf{\ }is obtained by $%
V_{1}\left( x,y\right) $ with $x,y$ interchanged. Therefore the Lie
symmetries admitted by the potential $V_{2}\left( x,y\right) $ are: 
\begin{equation*}
X=c_{0}\partial _{t}+c_{1}\sin t\partial _{x}\ +c_{2}\cos t\partial
_{x}+c_{3}x\partial _{x}
\end{equation*}%
and again in \cite{Leach Henon - Heiles problem} the Lie point symmetry $%
y\partial _{y}$ is missing.

The potential $V_{3}\left( x,y\right) $ is of the form $\frac{1}{2}\left(
x^{2}+y^{2}\right) +f\left( x-ay\right) $ and belongs to the types:

\begin{center}
\begin{tabular}{|l|l|l|l|l|}
\hline
\textbf{Table} & \textbf{Line} & \textbf{Parameters} & \textbf{Lie } & 
\textbf{Type of vector} \\ \hline
12 & 1 & $~m=-1,c_{1}=0,b=\pm 1,\neq 0,$ & ($\cos t$ or $\sin t)\left(
a\partial _{x}\pm \partial _{y}\right) $ & gradient KV \\ \hline
10 & 5 & $d=0,c_{1}=\frac{1}{2},f=\pm \left( ay\pm x\right) ^{3}$ & $\left(
ax+y\right) \left( a\partial _{x}+\partial _{y}\right) $ & two gradient KVs
\\ \hline
\end{tabular}
\end{center}

Therefore its Lie point symmetries are: 
\begin{equation*}
X=c_{0}\partial _{t}+(c_{1}\cos t+c_{2}\sin t)\left( a\partial _{x}\pm
\partial _{y}\right) +c_{3}\left( ax+y\right) \left( a\partial _{x}+\partial
_{y}\right).
\end{equation*}

The potential $V_{4}\left( x,y\right) $ is of the same form as $V_{3}\left(
x,y\right) $ with $x,y$ interchanged. Therefore the Lie point symmetries
are: 
\begin{equation*}
X=c_{0}\partial _{t}+a(c_{1}\cos t+c_{2}\sin t)\left( a\partial _{x}\mp
\partial _{y}\right) +c_{3}\left( ax+y\right) \left( a\partial _{x}+\partial
_{y}\right).
\end{equation*}

We observe that in all four cases the Lie symmetries depend on four free
parameters (the $c_{0},c_{1},c\,_{2},c_{3}).$ The parameter $c_{0}$
determines the vector $c_{0}\partial _{t}$ and the rest $c_{1},c\,_{2},c_{3}$
the $x-y$ part of the symmetry generators.

The Lie point symmetries which are possibly Noether symmetries are the ones
generated by the KVs. We check that the Lie point symmetries which are due
to the gradient KVs $\left( \text{with }m\neq 0\right) $ are Noether
Symmetries of the potentials (plus the $\partial _{\,t}$ whose Noether
integral is the Hamiltonian). The Noether integrals and the Noether
functions corresponding to each of these symmetries are given in Table 17.

\begin{center}
\bigskip

\begin{tabular}{lll}
\multicolumn{3}{l}{Table 17: Noether symmetries admitted by the potentials $%
V_{1},V_{2},V_{3},V_{4}$} \\ \hline
\multicolumn{1}{|l}{$\mathbf{V}\left( x,y\right) $} & \multicolumn{1}{|l}{%
\textbf{Noether Symmetry}} & \multicolumn{1}{|l|}{\textbf{Noether Integral}}
\\ \hline
\multicolumn{1}{|l}{$\frac{1}{2}\left( x^{2}+y^{2}\right) +x^{3}$} & 
\multicolumn{1}{|l}{$\sin t\partial _{y}\ $} & \multicolumn{1}{|l|}{$\dot{y}%
\sin t-y\cos t$} \\ \cline{2-3}
\multicolumn{1}{|l}{} & \multicolumn{1}{|l}{$\cos t\partial _{y}$} & 
\multicolumn{1}{|l|}{$\dot{y}\cos t+y\sin t$} \\ \hline
\multicolumn{1}{|l}{$\frac{1}{2}\left( x^{2}+y^{2}\right) +y^{3}$} & 
\multicolumn{1}{|l}{$\sin t\partial _{x}$} & \multicolumn{1}{|l|}{$\dot{x}%
\sin t-x\cos t$} \\ \cline{2-3}
\multicolumn{1}{|l}{} & \multicolumn{1}{|l}{$\cos t\partial _{x}$} & 
\multicolumn{1}{|l|}{$\dot{x}\cos t+x\sin t$} \\ \hline
\multicolumn{1}{|l}{$\frac{1}{2}\left( x^{2}+y^{2}\right) \pm \left( ay\pm
x\right) ^{3}$} & \multicolumn{1}{|l}{$\sin t\left( \mp a\partial
_{x}+\partial _{y}\right) $} & \multicolumn{1}{|l|}{$\left( \mp a\dot{x}+%
\dot{y}\right) \sin t-\left( \mp ax+y\right) \cos t$} \\ \cline{2-3}
\multicolumn{1}{|l}{} & \multicolumn{1}{|l}{$\cos t\left( \mp a\partial
_{x}+\partial _{y}\right) $} & \multicolumn{1}{|l|}{$\left( \mp a\dot{x}+%
\dot{y}\right) \cos t+\left( \mp ax+y\right) \sin t$} \\ \hline
\multicolumn{1}{|l}{$\frac{1}{2}\left( x^{2}+y^{2}\right) \pm \left( ay\mp
x\right) ^{3}$} & \multicolumn{1}{|l}{$\sin t\left( \pm a\partial
_{x}+\partial _{y}\right) $} & \multicolumn{1}{|l|}{$\left( \pm a\dot{x}+%
\dot{y}\right) \sin t-\left( \pm ax+y\right) \cos t$} \\ \cline{2-3}
\multicolumn{1}{|l}{} & \multicolumn{1}{|l}{$\cos t\left( \pm a\partial
_{x}+\partial _{y}\right) $} & \multicolumn{1}{|l|}{$\left( \pm a\dot{x}+%
\dot{y}\right) \cos t+\left( \pm ax+y\right) \sin t$} \\ \hline
\end{tabular}
\end{center}

These results coincide with those of \cite{Leach Henon - Heiles
problem,Shrauner}.

\subsection{Scalar field cosmology}

\label{Scalar field cosmology}

Scalar field cosmological models are used extensively in the study of
dynamics of the inflationary universe \cite{Kehagias Kofinas} and the
dynamics of unified UDM \cite{Bartacca Bartolo Matarrese,Vasilakos Lukes}.
In these models one normally considers a Friedmann Robertson Walker (FRW)\
space time and a scalar field with potential $V\left( \phi \right) $
minimally coupled to the gravitational field. The form of the scalar field
potential is taken at will, without a deeper physical or geometrical
hypothesis. In the following we propose that the form of the potential
should be fixed by the requirement that the Lagrangian of the equations of
"motion" admits a second Noether symmetry (besides the trivial $\partial
_{t} $). In this case one has two Noether symmetries, hence two Noether
integrals, and the dynamical system will be integrable.

The gravitational field equations\footnote{%
We set $8\pi G=c=1.$} of the FRW space time containing a scalar field $\phi $
only are (see e.g. \cite{Vasilakos Lukes}, \cite{Russo}):%
\begin{eqnarray}
3\left( H^{2}+\frac{k}{a^{2}}\right) &=&\frac{1}{2}\dot{\phi}^{2}+V\left(
\phi \right)  \label{SF.1} \\
2\frac{\ddot{a}}{a}+H^{2}+\frac{k}{a^{2}} &=&-\frac{1}{2}\dot{\phi}%
^{2}+V\left( \phi \right)  \label{SF.2}
\end{eqnarray}%
and are supplied by the field equation (known as the Klein Gordon equation)
for $\phi :$%
\begin{equation}
\ddot{\phi}+3H\phi +V^{\prime }(\phi )=0.  \label{SF.3}
\end{equation}%
Here $H=\frac{\dot{a}}{a}$ is the Hubble parameter, $k=0,\pm 1$ is the
curvature of the 3-d space and an overdot (respectively prime) indicates
derivation with respect to time (respectively $\phi ).$ The scalar field is
assumed to inherit\footnote{%
That is the scalar field satisfies the condition $\mathcal{L}_{\xi }\phi =0$
for all Killing vectors $\xi ^{a}$ of the metric. Because the 3-spaces are
maximally symmetric this implies that $\phi =\phi (t).$} the symmetries of
the space. Equation (\ref{SF.1}) is equivalent to the matter conservation
equation 
\begin{equation}
\dot{\mu}=-3(\mu +p)H  \label{SF.5}
\end{equation}%
where $\mu $ is the matter energy density of the FRW spacetime. This is
trivially satisfied, because we have assumed that space time contains a
scalar field only (hence $\mu =0$). We conclude that the scalar field
cosmological model (for all potentials $V(\phi )$!) can be considered as a
dynamical system with two degrees of freedom $a,\phi $ defined by the
equations (\ref{SF.2}), (\ref{SF.3}) or, equivalently, by the Lagrangian:%
\begin{equation}
L=3a\dot{a}^{2}-a^{3}\frac{\dot{\phi}^{2}}{2}+a^{3}V\left( \phi \right) -3ka.
\label{SF.4}
\end{equation}

We apply Theorem \ref{The Noether symmetries of a conservative system} to
determine the forms of the potential $V(\phi )$ for which the Lagrangian (%
\ref{SF.4}) admits Noether symmetries. To do that we write the Lagrangian in
the form \ $L=T-U$,~where:%
\begin{eqnarray}
T &=&3a\dot{a}^{2}-a^{3}\frac{\dot{\phi}^{2}}{2}  \label{SF.6} \\
U &=&-a^{3}V\left( \phi \right) +3ka.  \label{SF.7}
\end{eqnarray}%
In the phase space of $a,\phi $ the kinetic energy defines the metric:%
\begin{equation*}
ds^{2}=6ada^{2}-a^{3}d\phi ^{2}
\end{equation*}%
making this space an Einstein space. We compute the Ricci scalar $R=0,$
hence the space is a two dimensional Lorentzian flat space. Applying the
transformation:%
\begin{equation}
x=\sqrt{3}a^{3/2}\sinh \left( \sqrt{\frac{3}{2}}\phi \right) ,\;y=\sqrt{3}%
a^{3/2}\cosh \left( \sqrt{\frac{3}{2}}\phi \right)  \label{SF.8}
\end{equation}%
the metric is written in its canonical form and in the canonical coordinates 
$x,y$ the Lagrangian becomes:%
\begin{equation}
L=\frac{1}{2}\left( \dot{y}^{2}-\dot{x}^{2}\right) +\frac{1}{2}%
(y^{2}-x^{2})V\left( \frac{y}{x}\right) -k(y^{2}-x^{2})^{\frac{1}{3}}
\label{SF.009}
\end{equation}%
where we have absorbed the constants into $V\left( \frac{y}{x}\right) .~$

We determine the Lie and the Noether symmetries of this Lagrangian for the
cases $k=0~,~k\neq 0$ assuming that $V\left( \phi \right) \neq $constant.

Case 1: $k=0$

\noindent Lie symmetries

From Table 8, Line 4 with $d=0$ we find that the Lie symmetries are%
\begin{equation*}
X_{A}=c_{1}\partial _{t}+c_{2}\left( x\partial _{x}+y\partial _{y}\right)
\end{equation*}%
where $V\left( \frac{y}{x}\right) $ is an arbitrary function of its argument.

From Table 10, line 5 with $d\neq 0$ we find the Lie symmetry:%
\begin{equation*}
X_{B}=\left( c_{1}+c_{3}2t\right) \partial _{t}+c_{2}\left( x\partial
_{x}+y\partial _{y}\right) +c_{3}\frac{4}{d}\left( y\partial _{x}+x\partial
_{y}\right) 
\end{equation*}%
and $V\left( \frac{y}{x}\right) =e^{-d\arctan \frac{y}{x}}$. In the special
case $d=\pm 2$ then the system admits the additional Lie symmetry~$\partial
_{x}\pm \partial _{y}$

For $V(\frac{y}{x})=\frac{\omega _{1}}{2}\frac{x^{2}}{x^{2}-y^{2}}-\frac{%
\omega _{2}}{2}\frac{y^{2}}{x^{2}-y^{2}}$\ where $\omega _{1},\omega _{2}$\
are constants and $\omega _{1}\neq \omega _{2},~$the Lagrangian (\ref{SF.009}%
) describes the anisotropic oscillator.

From Table 8, lines 5 and 6 and Table 9 lines 1,2 we read the generic Lie
symmetry:%
\begin{equation*}
X_{C}=c_{1}\partial _{t}+\left( c_{2}\sin \left( \sqrt{\omega _{1}}t\right)
+c_{3}\cos \left( \sqrt{\omega _{1}}t\right) +c_{4}x\right) \partial
_{x}+\left( c_{5}\sin \left( \sqrt{\omega _{2}}t\right) +c_{6}\cos \left( 
\sqrt{\omega _{2}}t\right) +c_{7}y\right) \partial _{y}.
\end{equation*}

For $V(\frac{y}{x})=\frac{\omega _{1}}{2}\frac{x^{2}}{x^{2}-y^{2}}$ (i.e. $%
\omega _{2}=0$) we find the Lie symmetry:%
\begin{equation*}
X_{C_{1}}=c_{1}\partial _{t}+\left( c_{2}\sin \left( \sqrt{\omega _{1}}%
t\right) +c_{3}\cos \left( \sqrt{\omega _{1}}t\right) +c_{4}x\right)
\partial _{x}+\left( c_{5}+c_{6}t+c_{7}y\right) \partial _{y}.
\end{equation*}

For $V(\frac{y}{x})=-\frac{\omega _{2}}{2}\frac{y^{2}}{x^{2}-y^{2}}~$ (i.e. $%
\omega _{1}=0$) we find the Lie symmetry:%
\begin{equation*}
X_{C_{2}}=c_{1}\partial _{t}+\left( c_{2}+c_{3}t+c_{4}x\right) \partial
_{x}+\left( c_{5}\sin \left( \sqrt{\omega _{2}}t\right) +c_{6}\cos \left( 
\sqrt{\omega _{2}}t\right) +c_{7}y\right) \partial _{y}.
\end{equation*}

\noindent Noether symmetries

For $V\left( \frac{y}{x}\right) =$arbitrary there is only the standard
Noether symmetry $\partial _{t},$ with Noether integral the Hamiltonian $E$.

For $V\left( \frac{y}{x}\right) =e^{-d\arctan \frac{y}{x}}$ we have the
extra Noether symmetry from Table 14, line 3:%
\begin{equation*}
2t\partial _{t}+\left( x\partial _{x}+\frac{4}{d}y\partial _{x}\right)
+\left( y\partial _{y}+\frac{4}{d}x\partial _{y}\right) 
\end{equation*}%
with Noether integral%
\begin{equation*}
\phi =2tE+\left( x+\frac{4}{d}y\right) \dot{x}-\left( y+\frac{4}{d}x\right) 
\dot{y}.~
\end{equation*}

For the value $d=\pm 2,$ we have the extra Noether symmetry $\partial
_{x}\pm \partial _{y}~$whose Noether Integral is $\phi _{d=2}=\dot{x}\mp 
\dot{y}$\qquad 

For $V(\frac{y}{x})=\frac{\omega _{1}}{2}\frac{x^{2}}{x^{2}-y^{2}}-\frac{%
\omega _{2}}{2}\frac{y^{2}}{x^{2}-y^{2}}~$ we have the extra Noether
symmetries from Table 15 lines 1 and 2 
\begin{equation*}
X_{N}=\left( n_{2}\sin \left( \sqrt{\omega _{1}}t\right) +n_{3}\cos \left( 
\sqrt{\omega _{1}}t\right) \right) \partial _{x}+\left( n_{4}\sin \left( 
\sqrt{\omega _{2}}t\right) +n_{5}\cos \left( \sqrt{\omega _{2}}t\right)
\right) \partial _{y}
\end{equation*}%
with Noether integrals%
\begin{eqnarray*}
I_{n_{2}} &=&\sin \left( \sqrt{\omega _{1}}t\right) \dot{x}-\sqrt{\omega _{1}%
}\cos \left( \sqrt{\omega _{1}}t\right) x \\
I_{n_{3}} &=&\cos \left( \sqrt{\omega _{1}}t\right) \dot{x}+\sqrt{\omega _{1}%
}\sin \left( \sqrt{\omega _{1}}t\right) x \\
I_{n_{4}} &=&\sin \left( \sqrt{\omega _{2}}t\right) \dot{y}-\sqrt{\omega _{2}%
}\cos \left( \sqrt{\omega _{2}}t\right) y \\
I_{n_{5}} &=&\cos \left( \sqrt{\omega _{2}}t\right) \dot{y}+\sqrt{\omega _{2}%
}\sin \left( \sqrt{\omega _{2}}t\right) y
\end{eqnarray*}

For $V(\frac{y}{x})=\frac{\omega _{1}}{2}\frac{x^{2}}{x^{2}-y^{2}}~$we have,
from the same table and lines, the extra Noether symmetries%
\begin{equation*}
X_{N_{1}}=\left( n_{HO_{1}}\sin \left( \sqrt{\omega _{1}}t\right)
+n_{HO_{2}}\cos \left( \sqrt{\omega _{1}}t\right) \right) \partial
_{x}+\left( n_{F_{1}}+n_{F_{2}}t\right) \partial _{y}
\end{equation*}%
with Noether integrals%
\begin{eqnarray*}
I_{HO_{1}} &=&\sin \left( \sqrt{\omega _{1}}t\right) \dot{x}-\sqrt{\omega
_{1}}\cos \left( \sqrt{\omega _{1}}t\right) x \\
I_{HO_{1}} &=&\cos \left( \sqrt{\omega _{1}}t\right) \dot{x}+\sqrt{\omega
_{1}}\sin \left( \sqrt{\omega _{1}}t\right) x \\
I_{F_{1}} &=&\dot{y}~,~I_{F_{2}}=t\dot{y}-y.
\end{eqnarray*}

For $V(\frac{y}{x})=-\frac{\omega _{2}}{2}\frac{y^{2}}{x^{2}-y^{2}}~$we
have, from the same table and lines, the extra Noether symmetries%
\begin{equation*}
X_{N_{2}}=\left( n_{F_{1}}+n_{F_{2}}t\right) \partial _{x}+\left(
n_{HO_{1}}\sin \left( \sqrt{\omega _{2}}t\right) +n_{HO_{2}}\cos \left( 
\sqrt{\omega _{2}}t\right) \right) \partial _{y}
\end{equation*}%
with Noether integrals%
\begin{eqnarray*}
I_{HO_{1}} &=&\sin \left( \sqrt{\omega _{2}}t\right) \dot{y}-\sqrt{\omega
_{2}}\cos \left( \sqrt{\omega _{2}}t\right) y \\
I_{HO_{1}} &=&\cos \left( \sqrt{\omega _{2}}t\right) \dot{y}+\sqrt{\omega
_{2}}\sin \left( \sqrt{\omega _{2}}t\right) y \\
I_{F_{1}} &=&\dot{x}~,~I_{F_{2}}=t\dot{x}-x.
\end{eqnarray*}

\noindent Case 2: $k\neq 0\,\ $

\noindent Lie symmetries

For the arbitrary potential $V\left( \frac{y}{x}\right) $ there is only the
trivial Lie symmetry $\partial _{t}$.

For the potential $V\left( \frac{y}{x}\right) =e^{-C\arctan \frac{y}{x}}$ we
have from Table 10, line 5 with $d\neq 0$ the extra Lie symmetry: 
\begin{equation*}
\frac{2}{3}t\partial _{t}+\left( x\partial _{x}+y\partial _{y}\right) +\frac{%
4}{3C}\left( y\partial _{x}+x\partial _{y}\right) .
\end{equation*}

\noindent Noether symmetries

For all potentials only the trivial Noether symmetry $\partial _{t}$ is
admitted.

We conclude \ that in a flat (i.e. $k=0$) FRW spacetime with no matter
present the scalar field cosmological models which admit Noether symmetries,
hence possibly integrable, are the ones defined by the UDM potential $\left(
\omega _{1,2}<0\right) $ \cite{Vasilakos Lukes} and the exponential
potential. Indeed, the analytical solution for the exponential potential has
been given in \cite{Russo}.

\section{Conclusion}

\label{Conclusions}

We have presented two theorems which relate the Lie point symmetries and the
Noether symmetries of the second order system (\ref{L2P.1}) with the
generators of the special projective algebra and the homothetic algebra of
the metric respectively. Therefore if one knows these algebras of a
Riemannian space, then one computes the Lie (and the Noether) symmetries of
all dynamical systems moving in that space, by simply checking which $%
Y^{i\prime }s$ satisfy the appropriate constraint conditions with $F^{i}$.
We have applied these theorems to classify all two dimensional Newtonian
dynamical systems which admit at least one Lie point symmetry, and in the
case of conservative forces, all potentials $V(x,y)$ which admit a Lie point
symmetry and a Noether symmetry.

We have demonstrated the application of the results in various important
cases. We considered the Kepler-Ermakov system, which is an autonomous, but
in general not conservative dynamical system and we determined the classes
of this type of systems which admit Lie point and Noether symmetries; we
also considered the case of the H\`{e}non Heiles type potentials and
determined their Lie point symmetries and their Noether symmetries. These
results are compatible and complete previous results in the literature. We
continued with scalar field cosmological models and proved that the
exponential potential and the UDM potential are the only scalar field
potentials in a flat FRW\ spacetime in the absence matter, which lead to
integrable models.

The conclusion from the above is that the symmetries of a space modulate the
motion of dynamical systems moving in that space. That is, the space (i.e.
the metric)\ is not a simple substratum in which a dynamical system develops
its motion according to an external force (cause), but it is the space
itself which, via the special projective algebra, selects those forces which
posses Lie point symmetries and Noether symmetries. An equivalent way to
state this conclusion is to say that \emph{\ a dynamical system moving in a
Riemannian space has conserved quantities, in the standard sense of first
integrals and/or Noether integrals, only if the force causing the motion is
compatible with the geometry (i.e. metric) of the space.}

There are many directions that the present work can be extended. The
immediate topic to consider is the classification of all three dimensional
Newtonian dynamical systems, conservative and non-conservative, which admit
Lie point and Noether symmetries \cite{Damianou Sophocleous (2004) Nonlinear
Dynamics}. However initial work has shown, that this topic involves a large
number of cases and must be considered separately. Perhaps the next step to
be taken is to depart form the easy case of flat spaces and consider motion
in curved Riemannian spaces. Following \cite{Sen}, it would also be
interesting to use the above results, and study the relation between the Lie
point / Noether symmetries and the integrability of second order systems. Of
course, a wealth of results will follow if the theorems presented in this
paper are generalized to the case of a time dependent potential, a problem
on which we are currently involved.

\paragraph{ Acknowledgements}

The authors thank Professor P.G.L. Leach for valuable comments and
corrections; also an anonymous referee for the useful remarks and references
he kindly provided.

\appendix

\section{Appendix}

In this appendix we discuss the relation between the Lie symmetries of the
harmonic oscillator and those of a free particle.

We consider first the one dimensional case.

The equation of motion of the damped, attractive or repulsive, forced, time
dependent harmonic oscillator is:%
\begin{equation}
\frac{d^{2}x}{dt^{2}}+\gamma \frac{dx}{dt}+\varepsilon \omega ^{2}(t)-f(t)=0
\label{AM.5}
\end{equation}
where $\gamma $ is a constant and $\varepsilon =+1$ for the attractive and $%
-1$ for the repulsive oscillator.

Theorem 3.4 (p. 961) of \cite{Aminova 2006} states that:\newline
The second order differential equation%
\begin{equation}
\frac{d^{2}x}{dt^{2}}+a(t,x)\left( \frac{dx}{dt}\right) ^{3}+b(t,x)\left( 
\frac{dx}{dt}\right) ^{2}+c(t,x)\frac{dx}{dt}+d(t,x)=0  \label{AM.1}
\end{equation}%
can be reduced by a change of variables $(t,x)\rightarrow (\bar{t},\bar{x})$
to the equation 
\begin{equation}
\frac{d^{2}\bar{x}}{d\bar{t}^{2}}=0  \label{AM.2}
\end{equation}%
if and only if the following conditions hold:%
\begin{align}
3\frac{\partial ^{2}d}{\partial x^{2}}-2\frac{\partial ^{2}c}{\partial
t\partial x}+3d\frac{\partial b}{\partial x}+3b\frac{\partial d}{\partial x}%
-2c\frac{\partial c}{\partial x}+\frac{\partial ^{2}b}{\partial t^{2}}+c%
\frac{\partial b}{\partial t}-6d\frac{\partial a}{\partial t}-3a\frac{%
\partial d}{\partial t}& =0  \label{AM.3} \\
\frac{\partial ^{2}c}{\partial x^{2}}-2\frac{\partial ^{2}b}{\partial
t\partial x}-b\frac{\partial c}{\partial x}+3d\frac{\partial a}{\partial x}%
+6a\frac{\partial d}{\partial x}+3\frac{\partial ^{2}a}{\partial t^{2}}+2b%
\frac{\partial b}{\partial t}-3c\frac{\partial a}{\partial t}-3a\frac{%
\partial c}{\partial t}& =0.  \label{AM.4}
\end{align}%
Comparing (\ref{AM.5}) with (\ref{AM.1}) we find:%
\begin{equation*}
a=b=0,~ c=\gamma ,~ d=\varepsilon \omega ^{2}(t)x-f(t).
\end{equation*}

It is easy to show that for these values of $a,b,c,d$ conditions (\ref{AM.3}%
), (\ref{AM.4}) are satisfied, therefore there exists a coordinate
transformation which transforms equation (\ref{AM.5}) to equation (\ref{AM.2}%
). Because under a change of variables the Lie symmetry algebra of the
equation does not change, we infer that the damped, attractive or repulsive,
forced, time dependent harmonic oscillator shares the same Lie symmetry
algebra (i.e. the $sl(3,R),$ with the free particle \cite{Lutzky,Leach
Repulsive Oscillator,Cervero}.

We continue with the $n$ dimensional case.

The isotropic time dependent $n$ dimensional harmonic oscillator is defined
by the equation \cite{Prince Eliezer}: 
\begin{equation}
\frac{d^{2}x^{i}}{dt^{2}}+\varepsilon \omega ^{2}(t)x^{i}=0.  \label{AM.8}
\end{equation}%
In order to check that (\ref{AM.8}) is reducible to the free particle
equation of motion (i..e. the geodesic equations) we use Theorem 3.2. (p.
960) of \cite{Aminova 2006} which states that:

The differential system%
\begin{equation}
\frac{d^{2}x^{i}}{dt^{2}}+a_{jk}(x,t)\frac{dx^{i}}{dt}\frac{dx^{j}}{dt}\frac{%
dx^{k}}{dt}+b_{jk}^{i}(x,t)\frac{dx^{j}}{dt}\frac{dx^{k}}{dt}+c_{j}^{i}(x,t)%
\frac{dx}{dt}^{j}+d^{i}(x,t)=0  \label{AM.6}
\end{equation}%
is reducible by a change of variables $(t,x^{i})\rightarrow (\bar{t},\bar{x}%
^{i})$ to the differential system%
\begin{equation}
\frac{d^{2}\bar{x}^{i}}{d\bar{t}^{2}}=0  \label{AM.7}
\end{equation}%
if, and only if, for $i>3$ the Weyl projective tensor in the space $%
\{x^{i},t\}$ of the associated projective connection with components%
\begin{eqnarray*}
\Pi _{jk}^{i} &=&b_{jk}^{i}-\frac{1}{n+1}(b_{jk}^{m}\delta
_{m}^{i}+b_{mk}^{m}\delta _{j}^{m}) \\
2\Pi _{tj}^{i} &=&c_{j}^{i}-\frac{1}{n+1}c_{m}^{m}\delta _{j}^{i} \\
\Pi _{jk}^{t} &=&-a_{jk} \\
\Pi _{tt}^{i} &=&d^{i} \\
\Pi _{tj}^{t} &=&-\frac{1}{n+1}b_{mj}^{m} \\
\Pi _{tt}^{t} &=&-\frac{1}{n+1}c_{m}^{m}
\end{eqnarray*}%
vanishes. The corresponding transformation involves $n(n+2)$ arbitrary
constants.

Comparing (\ref{AM.8}) with (\ref{AM.6}) we find:%
\begin{equation*}
a_{jk}=0,b_{jk}^{i}=0,c_{j}^{i}=0,d^{i}=\varepsilon \omega ^{2}(t)x^{i}.
\end{equation*}%
It follows that for these values of the coefficients $a_{jk},~ b_{jk}^{i},
~c_{j}^{i},~ d^{i}$ the only non vanishing component of the projective
connection is:%
\begin{equation*}
\Pi _{tt}^{i}=\varepsilon \omega ^{2}(t)x^{i}.
\end{equation*}%
A standard computation shows that the Weyl projective tensor vanishes. Hence
the time dependent $n$ dimensional harmonic oscillator is equivalent to the
free particle moving in the $n$ dimensional space whose Lie symmetry algebra
is \cite{Prince Eliezer} the $sl(n+2,R)$. We note that the result holds for
any function $\omega ^{2}(t)$ hence we arrive at Leach's conclusion that the
Lie point symmetry algebra for any linear system in $n-$dimensions is the $%
sl(n+2,R)$.


\begin{thebibliography}{99}
\bibitem{Olver Book} Olver P J 1986 \emph{Application of Lie groups to
differential equations }(Springer Graduate texts in Mathematics, New York:
Springer) 

\bibitem{Stephani book ODES} Stephani H 1989 \emph{Differential Equations:
Their Solutions using Symmetry }(Cambridge University Press)

\bibitem{Aminova 2006} Aminova A V and Aminov N A 2006 Sbornic Mathematics 
\textbf{197} 951

\bibitem{Aminova2010} Aminova A V and Aminov N A 2010 Sbornic Mathematics 
\textbf{201} 631

\bibitem{Feroze Mahomed Qadir} Feroze T, Mahomed F M and Qadir A 2006
Nonlinear Dynamics \textbf{45} 65

\bibitem{Tsamparlis  Paliathanasis Lie geodesics Nonlinear Dynamics  2010} %
Tsamparlis M and Paliathanasis A 2010 Nonlinear Dynamics \textbf{62} 203

\bibitem{Fredericks Mahomed Qadir 2007} Fredericks E, Mahomed F M, Momoniat
E and Qadir A 2008 Computer Physics Comm. \textbf{179} 438

\bibitem{Sen} Sen T 1987\emph{\ }Phys. Lett. A \textbf{122} 327

\bibitem{Damianou Sophocleous 1999} Damianou P A and Sophocleous C 1999 
\emph{\ }J. Math. Phys. \textbf{40} 210

\bibitem{Karasu} Karasu A and Yildirim H 2002\emph{\ }J. Nonlinear Math.
Phys.\textbf{\ 9} 475

\bibitem{LeachK} Leach \ P G L and Karasu A 2004 J. Nonlinear Math. Phys. 
\textbf{11} 269

\bibitem{MoyoL} Moyo S and Leach P G L 2002 J. Phys. A \textbf{35} 5333

\bibitem{Leach Henon - Heiles problem} Leach P G L 1981\emph{\ \ }J. Math.
Phys. \textbf{22} 679

\bibitem{Shrauner} Abraham - Shrauner B 1990 J. Math. Phys. \textbf{31} 1627

\bibitem{Athorne 1991} Athorne C 1991 J. Phys. A \textbf{24} 1385

\bibitem{Kehagias Kofinas} Kehagias A and Kofinas G 2004 Class. Quantum
Grav. \textbf{21} 3871

\bibitem{Bartacca Bartolo Matarrese} Bertacca D, Matarrese S and Pietroni M
2007 Mod. Phys. Let. A \textbf{22} 38 2893

\bibitem{Vasilakos Lukes} Basilakos S and Lukes - Gerakopoulos G 2008 Phys.
Rev. D \textbf{78} 1550

\bibitem{Russo} Russo J G 2004 Phys. Lett. B \textbf{600} 185

\bibitem{Damianou Sophocleous (2004) Nonlinear Dynamics} Damianou P A and
Sofokleous C 2004\emph{\ }Nonlinear Dynamics \textbf{36} 3

\bibitem{Lutzky} Lutzky M 1978 J. Phys. A \textbf{11} 249

\bibitem{Leach Repulsive Oscillator} Leach P G L \ 1980 J. Phys. A \textbf{13%
} 1991

\bibitem{Cervero} Cervero J M and Villarroel J 1984 J. Phys. A \textbf{17}
1777

\bibitem{Prince Eliezer} Prince G E and Eliezer C J 1980 J. Phys. A \textbf{%
13} 815
\end{thebibliography}
\end{document}